\documentclass[%
 reprint,
 showpacs,
 amsmath,amssymb,
]{revtex4-1}
\usepackage{graphicx}
\usepackage{dcolumn}
\usepackage{bm}
\usepackage[mathlines]{lineno}

\usepackage[backref=none]{hyperref} 
\usepackage[usenames,dvipsnames]{color}
\definecolor{MyDarkBlue}{rgb}{0,0.1,0.75}
\hypersetup{pdfborder={0 0 0},colorlinks,breaklinks=true,
  urlcolor={MyDarkBlue},citecolor={MyDarkBlue},linkcolor={MyDarkBlue}
}

\begin{document}


\title{Polar morphologies from first principles: PbTiO$_3$ films on SrTiO$_3$ substrates and the $p(2 \times \Lambda)$ surface reconstruction}

\author{Jack S. Baker$^{1, 2}$}
\author{David Bowler$^{1, 2, 3}$}

\affiliation{$^1$London Centre for Nanotechnology, UCL, 17-19 Gordon St, London WC1H 0AH, UK \\ $^2$Department of Physics \& Astronomy, UCL, Gower St, London WC1E 6BT, UK \\
$^3$International Centre for Materials Nanoarchitectonics (MANA)
National Institute for Materials Science (NIMS), 1-1 Namiki, Tsukuba, Ibaraki 305-0044, Japan}%

\date{\today}

\begin{abstract}
Low dimensional structures comprised of ferroelectric (FE) PbTiO$_3$ (PTO) and quantum paraelectric SrTiO$_3$ (STO) are hosts to complex polarization textures such as polar waves, flux-closure domains and polar skyrmion phases. Density functional theory (DFT) simulations can provide insight into this order, but, are limited by the computational effort needed to simulate the thousands of required atoms. To relieve this issue, we use the novel multi-site support function (MSSF) method within DFT to reduce the solution time for the electronic groundstate whilst preserving high accuracy. Using MSSFs, we simulate thin PTO films on STO substrates with system sizes $>2000$ atoms. In the ultrathin limit, the polar wave texture with cylindrical chiral bubbles emerges as an intermediate phase between full flux closure domains and in-plane polarization. This is driven by an internal bias field born of the compositionally broken inversion symmetry in the [001] direction. Since the exact nature of this bias field depends sensitively on the film boundary conditions, this informs a new principle of design for manipulating chiral order on the nanoscale through the careful choice of substrate, surface termination or use of overlayers. Antiferrodistortive (AFD) order locally interacts with these polar textures giving rise to strong FE/AFD coupling at the PbO terminated surface driving a $p(2 \times \Lambda)$ surface reconstruction. This offers another pathway for the local control of ferroelectricity.

\end{abstract}
\maketitle


\section{\label{intro:level1} Introduction}

With the advent of advanced deposition techniques \cite{Schlom2008, Willmott2004} has come a revolution in the engineering of thin film perovskite oxides and layered  heterostructures for a variety of applications in nanoelectronics. These advancements have propelled research into the great variety of physical phenomena such systems can exhibit. These include enhanced colossal magnetoresistance \cite{Kobayashi1998}, high-temperature superconductivity \cite{Sleight1993}, the formation of interfacial two-dimensional electron and hole gases (2DEG/HG) \cite{Hwang2012, Yin2015} and the emergence of negative capacitance \cite{zubko2016negative}. While these are becoming well-documented \cite{Dawber2005, Mannhart2010, Zubko2011}, new emergent phenomena resulting from exotic electrical polarization textures, including polar waves, vortices and polar skyrmion phases \cite{Lu2018, Sichuga2011, Damodaran2017, Hong2018, Du2019} are less well understood. The toroidal moment born from the chirality of these polar morphologies can give rise to strong electrotoroidic, pyrotoroidic and piezotoroidic effects \cite{Prosandeev2008, Chen2015, Damodaran2017} all of which show promise to be exploited in new low dimensional functional devices. It is the purpose of this work to investigate these polar morphologies in low dimensional ferroelectric (FE) PbTiO$_3$ (PTO) on SrTiO$_3$ (STO) substrates using state-of-the-art methods based on density functional theory (DFT) designed for demanding calculations with large numbers of atoms \cite{Nakata2020large, Nakata2014}. \par 

Further complex phenomena can arise by the interacting order parameters of the system. Notably, many perovskites (and heterostructures) are susceptible to both the antiferrodistortive (AFD) and FE distortions. In the bulk, these two modes were thought to suppress one another, although, recent evidence suggests that a cooperative regime may also exist \cite{Gu2018, Aschauer2014}. At surfaces and interfaces we see phase coexistence. For example, at the PbO terminated [001] surface of PTO, we observe the AFD c($2 \times 2$) surface reconstruction. This is characterised by strong antiphase rotations of the TiO$_6$ octahedra about the [001] axis (or $a^0a^0c^-$ in Glazer's notation \cite{Glazer1972, Glazer1975}) and is known to coexist with and mutually enhance in-plane ferroelectricity \cite{Munkholm2001, Bungaro2005, Sepliarsky2005}. It is now popular to interface PTO with STO in the repeating (STO$_n$/PTO$_n$)$_{N}$ superlattice for $n$ alternating perovskite unit cells repeated $N$ times in a layered heterostructure. In the case of an ultrashort period ($n = 1$), hybrid-improper ferroelectricity can arise from the coupling of AFD and FE modes. \cite{Bousquet2008}. \par

 The (STO$_n$/PTO$_n$)$_{N}$ heterostructure continues to be studied from a theoretical perspective. Most frequently considered is full periodic boundary conditions with infinitely repeated layers. This has been studied with first principles DFT \cite{AguadoPuente2011} where the focus has been on the interactions of FE and AFD instabilities in the monodomain configurations. More recently, however, polydomain configurations have been studied and ferroelectric flux-closure domains have been shown to be stable \cite{AguadoPuente2012}. Since these simulations are of the bulk superlattice, there can be no account for the $c(2 \times 2)$ surface reconstruction. Polydomain simulations of pure TiO$_2$ terminated PTO films (which do not give rise to enhanced surface AFD modes \cite{Bungaro2005, Munkholm2001}) \textit{have} been performed but neglect the STO substrate by instead choosing to adopt the fictitious free-standing film geometry \cite{Shimada2010}. This work then cannot make any account for substrate/interface effects or the intrinsically broken inversion symmetry for epitaxially deposited films on substrates. To the author's knowledge, there has been only a single DFT-based study treating both the surface and the STO substrate. This study probed the nature of the theoretically proposed 2DEG/HG pair \cite{Yin2015} where thicker films ($\geq 14$ unit cells) of the FE monodomain out-of-plane configuration were considered as large depolarising fields are known to suppress this configuration in thinner films \cite{Junquera2003}. Multiple FE domains are considered a competing phase but no 2DEG/HG pair emerges as the resulting flux closure domains are an alternate mechanism for screening the depolarising field \cite{aguado2015model}. \par
 
 The polydomain configuration is challenging to simulate from the perspective of DFT. The difficulty arises from the increased number of atoms $N$ in the supercell as a result of treating a fixed domain period $\Lambda$. If we are also to treat a finite thickness of film $t$, the Kittel scaling law requires that the equilibrium $\Lambda$ too increases since $\Lambda \propto \sqrt{t}$ \cite{Streiffer2002, Fong2004}. We must also account for a significant amount of substrate as the large ferroelectric polarization of PTO may be able polarise a few layers of STO. Finally, if we wish to include AFD modes in our simulations (like the $c(2 \times 2$) surface reconstruction), the periodicity in the [010] direction must be doubled as to not frustrate the octahedral rotations. To include all of all of these effects, calculations with a supercell containing \textit{a few thousand atoms} are needed rather than the few hundred that conventional planewave DFT is able to handle \cite{bowler2006recent, bowler2012methods}. To overcome these limitations, some resort to alternative methods including phase fields, shell models, Monte Carlo and second principles \cite{ghosez2006first, garcia2016second, chen2002phase, Shafer2018, Damodaran2017, Sichuga2011, Chapman2017, Sepliarsky2006}. Whilst these approaches each have their merits, they cannot \textit{universally} serve as a replacement for full DFT. Most of these methods implicitly accept that full DFT offers superior accuracy since they either are or can be parameterised by the theory \cite{garcia2016second, Chapman2017, Sepliarsky2006}. Further, all of these methods make no account (or little account \cite{garcia2016second}) for the electronic structure, limiting the ability of simulations to adapt to new chemical environments like at surfaces and interfaces with a substrate.  In this work, we utilise a novel variation on full local orbital DFT which allows us to consider systems of a few thousand atoms whilst preserving high chemical accuracy \cite{Nakata2014}. \par
 
 We provide a full first principles study on ultrathin films of PTO down to a single PbO monolayer ($N_z = 0$) up to 9 unit cells ($N_z = 9$) where $N_z$ is the number of perovskite unit cells stacked in the [001] direction. We treat explicitly the STO substrate and the PbO termination invoking the c($2 \times 2$) surface reconstruction. As well as treating paralelectric films, we treat two monodomain configurations of the polarization ($\mathbf{P}$ $||$ [100] and $\mathbf{P}$ $||$ [110]) and stripe domain patterns comprised of alternating $\mathbf{P}$ $||$ [001] and $\mathbf{P}$ $||$ [$00\bar{1}$] domains. We do not consider monodomain polarization oriented in the out-of-plane directions ($\mathbf{P}$ $||$ [001] or $\mathbf{P}$ $||$ [00$\bar{1}$]) since a previous work has shown this is suppressed by the depolarising field within our range of thicknesses (also verified to be true within our simulations). We characterise the competing energetics of the different film geometries as well detailing the resulting polar morphologies, interactions with the surface reconstruction and other structural features. This work is then organised as follows: section \ref{mssf:level2} constitutes an overview of the multi-site support function method as applied in this study whilst Section \ref{calcdetails:level2} presents finer details associated with the simulation method. Section \ref{supercell:level2} details the treated film geometries for each thickness of film. In section \ref{competphase:level2} we compare the competing energetics between each film geometry as a function of thickness. In section \ref{morph:level2}, the local polarization fields of the various film geometries are discussed and in Section \ref{surfrecon:level2}, we detail the amplitudes of local AFD modes, including the $c(2 \times 2)$ surface reconstruction, its coupling with surface polarization and the asymmetrical rumpling of surface/interfacial Pb cations. We conclude this work in Section \ref{summary:level1} providing a summary of the findings of this investigation, commenting on the impact this work has on the field and suggesting new areas for which our novel method is applicable. 

\section{\label{method:level1} Theoretical method}
 
Simulations based on DFT have provided accurate first-principles descriptions of countless condensed matter and molecular systems since the 1970s. The method is centered upon the self-consistent calculation of the charge density $n(\mathbf{r})$ for a system of interacting ions and electrons occupying independent Kohn-Sham \cite{Kohn1965} orbitals. The ground-state total energy of the system is evaluated as a uniquely defined functional of this $n(\mathbf{r})$ as detailed in the Hohenberg-Kohn theorem \cite{Hohenberg1964}. This method would allow for exact calculations if it weren't for the small electronic exchange and correlation terms for which the exact functional is unknown, thus, (increasingly accurate) approximations must be made. In this work, calculations were performed using the implementation of DFT present in the \textsc{CONQUEST} code \cite{Nakata2020large, CQRelease2020}. This code is designed for large-scale, massively parallel simulations on thousands (even millions \cite{Bowler2010}) of atoms, utilizing the latest in high performance computing (HPC). While we now focus on the multi-site support functions used in this work (Section \ref{mssf:level2}), we refer the reader to \cite{Nakata2020large, bowler2006recent, bowler2012methods} for further details on the \textsc{CONQUEST} methodology and to \cite{CQRelease2020} for a recent (freely available) public release of the code. 
    
    \subsection{\label{mssf:level2} Multi-site support functions}
    
    This section seeks to provide a convenient overview of the multi-site support function method (MSSF) method as implemented in \textsc{CONQUEST} \cite{Nakata2020large, CQRelease2020}. For a more detailed discussion, the reader is referred to \cite{Nakata2014, rayson2009highly} and references therein. 
    
    \par
    
     We begin by considering the Kohn-Sham equations in a non-orthogonal basis in terms of the Hamiltonian and overlap matrix elements $H^{\mathbf{k}}_{i\alpha j\beta}$ and $S^{\mathbf{k}}_{i\alpha j\beta}$ 
     
    \begin{equation}
       H^{\mathbf{k}}_{i\alpha j\beta} c_{j\beta}^{n\mathbf{k}} = \epsilon_{n\mathbf{k}} S^{\mathbf{k}}_{i\alpha j\beta} c_{j\beta}^{n\mathbf{k}}
        \label{KSEq}
    \end{equation}
    
     for ionic sites $i$, $j$ orbital indices $\alpha$, $\beta$, eigenvalues $\epsilon_{n\mathbf{k}}$ and wavevector $\mathbf{k}$. Here, the Kohn-Sham eigenstates are written as a linear combination of support functions $\phi_{i\alpha}(\mathbf{r})$ 
     
     \begin{equation}
         \psi_{n\mathbf{k}}(\mathbf{r}) = \sum_{i\alpha} c_{i \alpha}^{n \mathbf{k}} \phi_{i\alpha}(\mathbf{r})
         \label{SFComb}
     \end{equation}
     
     for support function coefficients $c_{i \alpha}^{n \mathbf{k}}$. A support function is a local orbital which we define to become zero at some cutoff radius $r_c$ ($\phi_{i \alpha}(\mathbf{r}) = 0$, $|\mathbf{r} - \mathbf{R}_i| \geq r_c$). It is then possible to represent these support functions in a compact, efficient basis of atomic-centered functions whose spatial cutoff leads to the onset of matrix sparsity (important to the $\mathcal{O}(N)$ mode of operation). How we choose to do so is paramount as this determines the accuracy and computational effort required for the calculation of the electronic groundstate. For accurate calculations we generally employ multiple-$\zeta$ basis-sets \cite{Bowler2019, Baker2020pseudoatomic, soler2002siesta}. The most accurate evaluation of the support function coefficients here comes from the mapping of one $\zeta$ to one support function and the subsequent diagonalisation of a \textit{large} $\mathbf{H^{k}}$. This, however, comes with the burden of the greatest computational cost since this operation scales cubically with the total number of basis functions. Another choice is to make a basis set expansion of $\phi_{i\alpha}(\mathbf{r})$   
    
    \begin{equation}
        \phi_{i\alpha}(\mathbf{r}) = \sum_{\zeta} d_{i\alpha, i\zeta}\chi_{i\zeta}(\mathbf{r})
        \label{LinearComb}
    \end{equation}
    
    where index $\chi_{i\zeta}$ is $\zeta$th atomic basis function on atomic site $i$. The coefficients $d$ can be found through an optimisation method such as the conjugate gradients scheme. Another choice for equation \ref{SFComb} is to build a support function from all of the basis functions on atomic site $i$. This is a simple extension of eq \ref{LinearComb} where instead of having $\zeta \in \alpha$, $\alpha$ spans all $\zeta$ on site $i$. While performing any of the previous two contractions does indeed reduce the number of required support functions, we are still unable to achieve a minimal representation due to atomic symmetry constraints \cite{torralba2008pseudo}.
    
    \par
    
    To achieve this minimal representation, we use the multi-site contraction
    
    \begin{equation}
        \phi_{i\alpha}(\mathbf{r}) = \sum_k^{\text{neighbours}} \sum_{\zeta \in k} C_{i\alpha, k\zeta}\chi_{k\zeta}(\mathbf{r})
        \label{MultiComb}
    \end{equation}
    
    where $k$ is an atomic site enclosed within a sphere of radius $r_{\text{MS}}$ about target atom $i$, inclusive of $i$. As the coefficients $C$ now span a subspace of a local molecular orbital (MO), we are now relinquished from the atomic symmetry constraints. To find the optimal balance between efficiency and accuracy, we use the local filter diagonalisation method \cite{rayson2009highly, rayson2010rapid} to evaluate the coefficients. This requires the solution of a local eigenproblem in a subset $s$ enclosed by a sphere of radius $r_{\text{LD}}$($\geq r_{\text{MS}}$) for each target atom.
    
    \begin{equation}
        \mathbf{H}_s \mathbf{C}_s = \mathbf{\epsilon}_s \mathbf{S}_s \mathbf{C}_s
        \label{LocalEig}
    \end{equation}
    
    For subspace eigenvalues $\mathbf{\epsilon}_s$, MO coefficients $\mathbf{C}_s$ and where $\mathbf{H}_s$ and $\mathbf{S}_s$ are the local subsets of the system-wide Hamiltonian and overlap matrices respectively. By projecting $\mathbf{C}_s$ onto a set of trial vectors $\mathbf{t}$, we obtain the contraction coefficients $\mathbf{C}'$
    
    \begin{equation}
        \mathbf{C}' = \mathbf{C}_s \mathbf{C}_s^T \mathbf{S}_s \mathbf{t}
        \label{Contract}
    \end{equation}
    
    Lastly, $\mathbf{C}^\prime$ is mapped onto the corresponding position in the contraction coefficient vectors, whose elements are extracted for use in eq \ref{MultiComb}. 
    
    \par
    
    The MSSF method therefore introduces two new adjustable parameters to the DFT calculation; $r_{\text{LD}}$ and $r_{\text{MS}}$. The former expands the space used for the local filter diagonalisation and the latter modifies the representation of $\phi_{i\alpha}(\mathbf{r})$. The accuracy of the contraction can be improved by increasing both of these subject to $r_{\text{MS}} \leq r_{\text{LD}}$. In practice, convergence can be achieved for relatively modest values of both (1-2 lattice constants of the bulk). Using this method, high accuracy calculations with a few thousand atoms can feasibly be performed on most standard HPC systems. Structural relaxations of $\approx 1000$ atoms (to a stringent force tolerance, as seen in this work) can be expected to take 1-2 weeks of wall-time on $\approx 200$ physical cores. Simulations of $\approx 2000$ atoms may need $\approx 500$ physical cores to meet the memory requirements, dependant on the memory-per-node on the HPC system.

    \subsection{\label{calcdetails:level2} Simulation details}
    
   Within the MSSF mode of operation described in section \ref{mssf:level2}, we find that $r_{\text{MS}} = r_{\text{LD}} = 6.35$\AA\ sufficiently converges relevant quantities (discussed in the supporting information). All simulations use the local density approximation (LDA) as parameterised by Ceperley, Alder, Perdew and Zunger \cite{perdew1981self, ceperley1980ground} to describe exchange and correlation effects. This functional has recently been shown to perform well against higher rung functionals for describing the properties of the perovskite oxides \cite{Zhang2017}. Although both hybrid and meta-GGA functionals better predict the experimental structural properties, they still overestimate the bulk polarization of PTO. Since this is the primary order parameter of our system, LDA is a good choice as the magnitude of the overestimate is much less. \par
   
   Optimised Vanderbilt norm-conserving pseudopotentials are used to replace core electrons. \cite{hamann2013optimized, Vanderbilt1990}. We use the scalar-relativistic variety available in the \texttt{PseudoDojo} library (\texttt{v0.4}) \cite{van2018pseudodojo}. These are used as an input for the \texttt{ONCVPSP} code (\texttt{v3.3.1}) \cite{hamann2013optimized}. We then generate a double-$\zeta$ plus double-polarization (DZDP) basis set of pseudo-atomic orbitals (PAOs) describing the Pb 5d 6s 6p 6d$^p$, Sr 4s 4p 5s 4d$^p$, Ti 3s 3p 4s 3d 4p$^p$ and the O 2s 2p 3d$^p$ orbitals respectively. Those orbitals superscripted with $p$ are polarization orbitals aimed at increasing the angular flexibility of the basis. Ti 3s, Ti 3p and Pb 5d orbitals are treated as semi-core states described only with a single $\zeta$. These PAOs are used to represent the support functions in section \ref{mssf:level2}. The PAO basis constructed using this set of pseudopotentials has been shown to provide high accuracy lattice constants, bulk moduli and electronic properties \cite{Bowler2019, Baker2020pseudoatomic}.
    
    \par
    
    Momentum space integrals are performed on a non-$\Gamma$-centered $6/N_x \times 6/N_y \times 1$ uniform mesh as described by Monkhorst \& Pack \cite{monkhorst1976special} where $N_{x, y}$ are the number of perovskite unit cells included in the [100] and [010] directions of the supercell respectively. While many real-space integrals are performed using intuitive analytic operations \cite{torralba2008pseudo}, some are performed on a fine, regular integration grid with a plane-wave equivalent cutoff of 300 Ha. This yields the cubic $Pm\bar{3}m$ lattice constants of PTO and STO ($a_{\text{PTO}}$ and $a_{\text{STO}}$) as 3.904$\text{\AA}$ and 3.874$\text{\AA}$ respectively. Further, the tetragonality (c/a) of FE $P4mm$ PTO is obtained as 1.04 producing a spontaneous polarization of 79.02 $\mu C/cm^2$ as calculated with Resta's method \cite{Resta1999,Resta1993} within the modern theory of polarization. This method is equivalent to the Berry phase \cite{Berry1984} formula of King-Smith and Vanderbilt \cite{KingSmith1993} in the limit of large cells; we use 10 PTO unit cells in the direction of the FE distortion to achieve convergence. Each of the structures in section \ref{supercell:level2} are fully relaxed (subject to constraints) with quenched molecular dynamics until the maximum absolute value of the force on every atom falls below $0.01$ eV/\AA. Since our slabs feature asymmetric surface terminations, a small dipole moment can develop in the out-of-plane direction. This propagates to a spurious electric field across the supercell which we correct using the scheme of Bengtsson \cite{bengtsson1999dipole}.
    
    \subsection{\label{supercell:level2} Supercell configurations}
    
\begin{figure}
       \includegraphics[width=\linewidth]{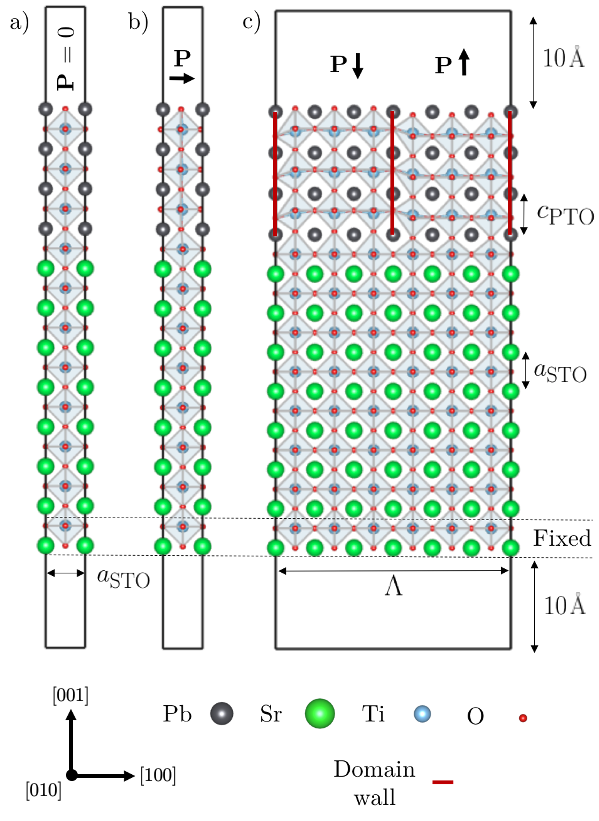}
        \caption{The initial supercell configurations for the $N_z = 3$ films before structural relaxation. Shown here are the supercells \textit{not} including AFD modes. Each configuration is however also treated with AFD modes following the explanation in section \ref{supercell:level2}.  a) The paraelectric supercell constrained such that spontaneous polarization cannot emerge. b) The monodomain in-plane ferroelectric case ($\mathbf{P}$ $||$ [100] is shown here, but we also treat $\mathbf{P}$ $||$ [110]) constrained such that spontaneous polarization cannot develop in the out-of-plane direction. c) The polydomain ferroelectric case with equally sized up and down domains for the ferroelectric polarization. Shown here is the $\Lambda = 6$ case.}
        \label{fig:InitStructures}
    \end{figure}

     The film geometries considered in this work are displayed in Figure \ref{fig:InitStructures}. Universal to all structures is a fixed amount of STO substrate. We find that 8/7 alternating SrO/TiO$_2$ monolayers is sufficient to converge the relaxed ionic positions of the PTO films (demonstrated in the supporting information, in agreement with a similar classical study \cite{Sepliarsky2006}). An interfacial region $a_I = 1/2(a_{\text{STO}} + c_{\text{PTO}})$ exists between the first PbO layer and last SrO layer which we assign to belong to neither the film or substrate. The in-plane components of the supercells are held to integer multiples of $a_{\text{STO}}$. The two bottom-most monolayers of the substrate have their atomic positions fixed in structural relaxations. These measures ensure we are applying a realistic mechanical strain to the PTO film as well as simulating the effects of a semi-infinite substrate. To limit unfavorable interactions with images of the film in the [001] direction, we introduce a total of $20\text{\AA}$ of vacuum. All supercells have the generalised dimensions \{$N_x a_{\text{STO}}$, $N_y a_{\text{STO}}$, $7a_{\text{STO}} + a_I + N_z c_{\text{PTO}} + 20\text{\AA}$\}. We simulate films of thickness $N_z = $ 0, 1, 2, 3, 5, 7 and 9 formula units of PTO where a thickness of 0 corresponds to a single PbO monolayer. Films of $N_z \geq 3$ unit cells were increased in steps of two unit cells such that the equilibrium domain period predicted by Kittel scaling could increase by an even integer number of unit cells (a requirement for domains to have an equal number of unit cells). This range of film thicknesss could encompass different energetically stable geometries including a transition between FE monodomain, polydomain and possible intermediate phases. It also spans low dimensional films with strong interface coupling with such effects decreasing with increased thickness.  

    \begin{figure}
       \includegraphics[width=\linewidth]{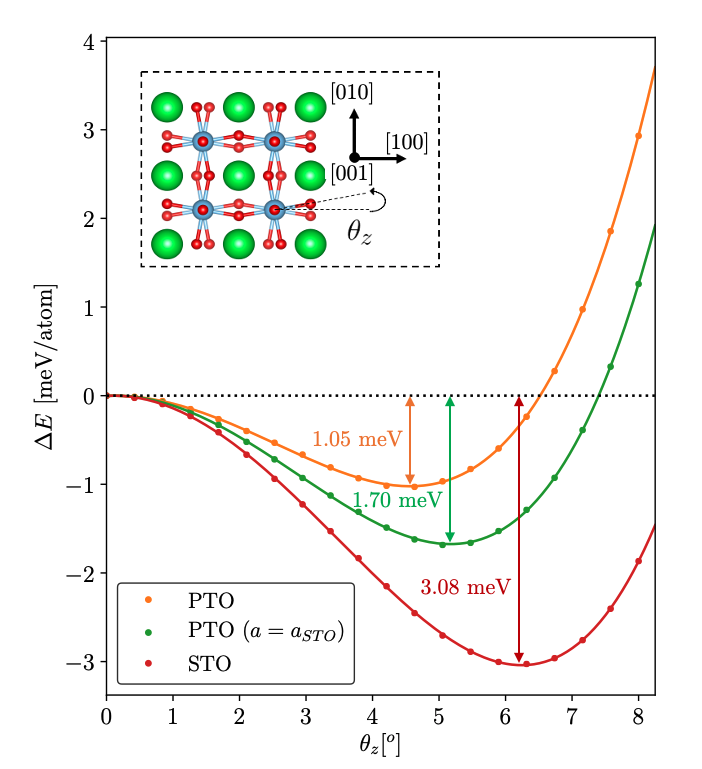}
        \caption{Energy versus the $R_4^+$ mode rotation angle $\theta_z$ for PTO at its bulk cubic lattice constants, strained PTO with $a=b=a_{\text{STO}}$ ($\sigma=-0.78\%$) and STO at its bulk cubic lattice constants. Calculations here are in full periodic boundary conditions in the infinite bulk crystal. The coefficients of quartic polynomials are evaluated with a least squares fitting procedure and plotted as lines. \textit{Inset} An illustration demonstrating the $R_4^+$ rotation angle $\theta_z$ shown here for STO. The same angle persists in PTO}
        \label{fig:BulkRotPlot}
    \end{figure}
    
    We choose to treat the following supercell configurations with \textit{and without} the influence of AFD modes. Note also that structures treated without these modes do not show the $c(2 \times 2)$ surface reconstruction. To do so, we must set $N_y = 2$ and have $N_x$ set to multiples of 2. The important AFD distortions for both STO and PTO are the zone-boundary $R_4^+$ and $M_3^+$ modes. The former is equivalent to a $a^0a^0c^-$ Glazer tilt system \cite{Glazer1972, Glazer1975} while the latter is $a^0a^0c^+$. The former has been found previously to be the most energetically favorable so we do \textit{not} treat the $M_3^+$ mode \cite{lebedev2009ab}. The $R_4^+$ mode can be defined with a single rotation angle $\theta_z$ as shown in the inset of Figure~\ref{fig:BulkRotPlot}. We find that the angles $\theta_z = 6.21^{\circ}$ and $\theta_z = 4.52^{\circ}$ (Figure \ref{fig:BulkRotPlot}) define minima in the energy of the cubic supercells of STO and PTO respectively. Since simulations are performed with the in-plane lattice constants of STO, we initialise supercells with the optimal strained PTO rotation angle $\theta_{z,\sigma} = 5.20^{\circ}$ as found in the tetragonal supercell. These results indicate that strains of $<1\%$ in PTO are able to both increase the optimal $\theta_z$ and increase the depth of the energy minima. These angles overestimate what is seen in experiment. For $I4/mcm$ STO,  $\theta_z \approx 2.1^{\circ}$ \cite{Lytle1964} while AFD modes are not observed in the PTO bulk. Calculations using hybrid functionals have been able improve this angle for STO ($\theta_z = 1.9^{\circ}$ \cite{heifets2006calculations}) but are not used in this study in part because of the computational cost but also due to the aforementioned overestimate in the bulk polarization of PTO.
    
        \subsubsection{\label{para:level3} Paraelectric}
        First we consider films of the high symmetry, paraelectric, non-polar variety  (Figure \ref{fig:InitStructures}a). These supercells are initialised by PTO formula units with the high symmetry cubic fractional atomic positions and optimal cubic out-of-plane lattice constant $c_{\text{PTO}}=3.904$\AA. Although spatial inversion symmetry is intrinsically broken by the composition of our supercells, we constrain atomic relaxations of these films to the lowest symmetry space group ($P4mm$) carefully preventing any cation-anion counter motion. This allows for the most degrees of freedom in relaxations without incurring any intrinsic ferroelectric polarization. It is the purpose of these films to provide a baseline for the relative stability of polar phases.
    
        \subsubsection{\label{inplane:level3} Monodomain in-plane ferroelectric}
        In this supercell, we allow monodomain ferroelectric polarization to develop in the [100] and [110] directions (Figure \ref{fig:InitStructures}b). PTO formula units retain the same $c_{\text{PTO}}$ as the paraelectric case but atomic positions now correspond to the $\Gamma_4^-$ mode of the primitive cubic PTO unit cell orientated in the [100] direction. This induces a ferroelectric polarization of 56.9 $\mu C/cm^2$ in the infinite bulk as calculated with the method described in Section \ref{calcdetails:level2}. To create [110] polarization, we simply displace the metal cations along the [110] direction and displace Oxygen anions along the [$\bar{1}\bar{1}0$] direction proportional to the magnitude of the bulk $\Gamma_4^-$ mode. During structural relaxation, we apply symmetry constraints to prevent the development of out-of-plane polarization \cite{Bungaro2005} although the non-trivial depolarising field that would result naturally suppresses it.
    
        \subsubsection{\label{poly:level3} Polydomain ferroelectric}
        Here we consider films intialised with a striped domain structure consisting of alternating regions of PTO polarised in the [001] and [00$\bar{1}$] directions which we will from this point onwards refer to as up and down domains respectively (Figure \ref{fig:InitStructures}c). Up and down domains are equal in size ($N_{\text{up}} = N_{\text{down}}$) and together form a full domain period $\Lambda$. Equivalently, $N_{\text{up}} + N_{\text{down}} = N_{x} \equiv \Lambda$. Domain walls are chosen to be centered on the PbO plane. This choice, however, is arbitrary as the energy difference between this and the TiO$_2$ centering is found to be $\sim 1$ meV per unit cell \cite{Shimada2010}. This is (slightly) beyond the resolution of our calculations, a fact noted in a comparable study. \cite{AguadoPuente2012}.
        
        \par
        
        It is found that there is a minimum thickness of film for this type of ferroelectricity to occur. Theoretical results of the free standing PTO film have shown that the polydomain ferroelectric film is lower in energy than the paraelectric configuration \cite{Shimada2010} but make no account of possible monodomain in-plane orientations for the polarization. In experiments conducted with an STO substrate \cite{Fong2004} it was found that that the polydomain configuration was only observed above 3 unit cells in thickness (in the temperature range of 311-644K). We also confirm that no out-of-plane component of $\mathbf{P}$ remains after structural relaxation of polydomain $N_z = 1$ and 2 films ($\Lambda = 4$). We then only consider this configuration for those film thicknesses $N_z \geq 3$. PTO unit cells are initialised with the strained FE $P4mm$ unit cell. When the in-plane constants are constrained to $a_{\text{STO}}$, this results in $c =4.049$\AA\ and a slightly enhanced polarization of 80.07 $\mu C/cm^2$.
        
        \par
        
        An accurate account for the polydomain structure requires us to work at the equilibrium domain period. To avoid the need to manually find this domain period for each thickness (which would require us to simulate several domain periods for each film thickness), we use knowledge from previous experimental and theoretical data. In particular, we find that both theory and experiment agree upon $\Lambda = 6$ when $N_z = 3$ \cite{Shimada2010, Fong2004}. We use X-ray diffraction data \cite{Streiffer2002} for the remaining film thicknesses. We choose the nearest even number of unit cells (to preserve the periodicity of the AFD rotation pattern) corresponding to the experimental domain periods. We then have for the $N_z = 3, 5, 7, 9$ films, domain periods of $\Lambda = 6, 8, 10, 12$ unit cells respectively. Films of this configuration are free from constraints during structural relaxation. 
        
        \par
        
        In a study of monodomain out-of-plane polarization, it was found that $\mathbf{P}$ $\parallel$ [00$\bar{1}$] (towards the STO substrate) becomes more polar than $\mathbf{P}$ $\parallel$ [001] (towards the vacuum). For polydomain films, this could mean that up domains are less polar than down domains. This would result in a small net dipole moment in the [00$\bar{1}$] direction (and a spurious electric field) developing during structural relaxation which is corrected for using the scheme described in Section \ref{calcdetails:level2}.

\section{\label{results:level1} Results}

    \subsection{\label{competphase:level2} Competing phases}
\begin{figure}
       \includegraphics[width=\linewidth]{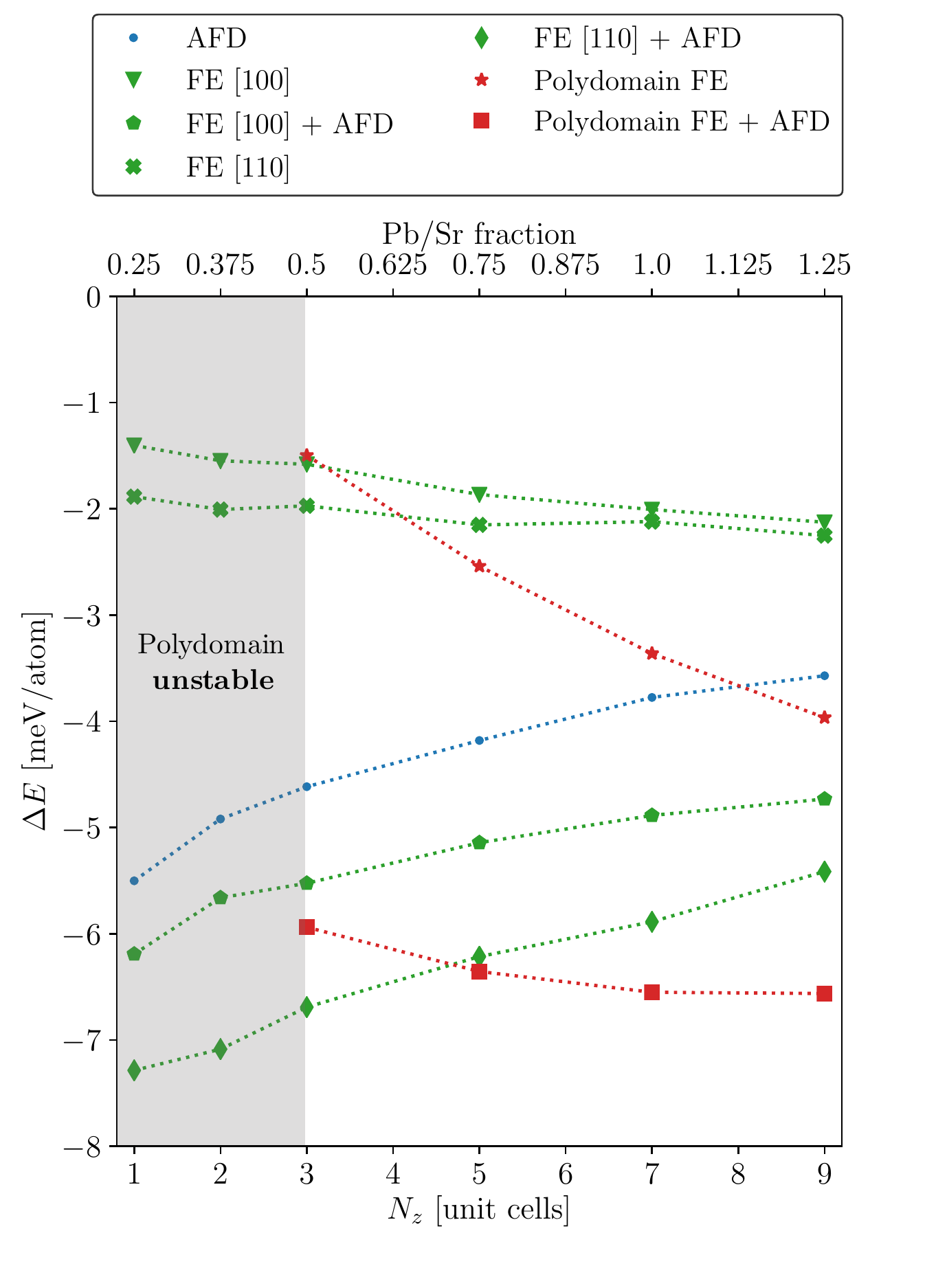}
        \caption{The energy difference $\Delta E$ compared with the non-polar paraelectric films for the film configurations considered in Figure \ref{fig:InitStructures} versus the film thickness in PTO unit cells, $N_z$. Since we use a fixed amount of STO substrate, the formula unit of the film alters with $N_z$. This is accounted for by the upper x-axis indicating the Pb/Sr fraction. The area in grey indicates the domain of $N_z$ for which the polydomain configuration is not considered.}
        \label{fig:Energetics}
    \end{figure}

    The film geometries considered have energetics which evolve as a function of film thickness. Figure \ref{fig:Energetics} displays this behaviour indicating the favorability of different phases. We measure this favorability as the energy difference $\Delta E$ between the geometry of film in question and the paraelectric film. We choose to measure $\Delta E$ in meV/atom since the (more common) definition of meV/formula unit would vary with film thickness. This is because as the film thickness increases, the Pb/Sr fraction (upper axis of Figure \ref{fig:Energetics}) increases as an artefact of a fixed amount of STO substrate. As a result, we expect to see a component energy lowering from ferroelectric phases as we increase $N_z$ since the energy of PTO is lowered with the onset of ferroelectricity (by 9.58 meV/atom in the bulk). We then expect to observe a rise in energy for purely AFD phases since the the fraction of STO, favoring AFD modes, has decreased. The energetics of film thickness $N_z$ = 0 are not present on Figure \ref{fig:Energetics} since no ferroelectric phase was stable. Adding AFD rotations does however lower the energy compared to the paraelectric film by an amount similar to $N_z$ = 1. \par  
    
    Considering first the monodomain in-plane ferroelectric films, Figure \ref{fig:Energetics} shows that the favored axis for the polarization is always [110] as was shown in a DFT study of the free standing film under compressive strain \cite{Umeno2006}. This is true with and without the influence of AFD modes. This favorable direction seems to diminish with film thickness becoming almost degenerate with [100] polarization at $N_z = 7$ for the films \textit{not} influenced by AFD modes. When AFD modes \textit{are} taken into account, the degree of favorability for [110] polarization (compared with [100]) almost doubles showing that [110] polarization is far more compatible with $a^0a^0c^-$ rotations. We suggest that [110] polarization is more favorable than [100] since $\mathbf{P} \parallel [100]$ is stunted by the epitaxial strain. An increased distortion along the longer diagonal axis of the supercell (of length $\sqrt{2}a_{\text{STO}}$) when compared with a distortion parallel to one of the pseudocubic axes (of length $a_{\text{STO}}$) relieves this stunting. We can also deduce whether coupling between AFD and FE modes is cooperative or competitive. The sum of $\Delta E$ for the FE [100] curve and the AFD curve is always lower than the combined FE [100] + AFD curve. This indicates that the the coupling is competitive with AFD modes suppressing FE ones and vice versa as is usually true for bulk modes. When making the same comparison for [110] polarization (which is \textit{not} observed in the bulk), however, it very closely mirrors the combined FE [110] + AFD curve. This suggests that FE and AFD modes are at worst independent of one another, but, for $N_z = 2$ or 3 are mildly cooperative, with the FE [110] + AFD curve being lower in energy than the sum by $\approx 0.2$ meV/atom (close to the resolution of the simulation). \par
    
    Remarkably, the polydomain configuration is not universally the ground state. Monodomain [110] polarization is the lowest in energy until a film thickness of 4-5 unit cells. This is close to the experimental observation of a polydomain structure at a thickness of 3 unit cells \cite{Fong2004} with the difference perhaps being an artefact of finite temperature in experiment. We note also that these results agree with the theoretical findings of Shimada \cite{Shimada2010} in that at a thickness of 3 unit cells, the polydomain configuration is lower in energy the the paraelectric film. In this work, however, while the energy is lowered by this geometry, monodomain in-plane [110] polarization is favored at this depth. It is important to note that this work treats the PbO termination whilst the work of Shimada treats the TiO$_2$ termination. As we discuss later (in Section \ref{morph:level2}), the $N_z = 3$ polydomain film \textit{does not} condense the flux-closure domain morphology but instead shows a polar wave (Figure \ref{fig:VecField}b). This clearly will have an impact on the energetics. Comparing the energy of the combined polydomain FE + AFD curve with the sum of the polydomain FE and AFD curves, we find that the latter is always lower in energy (a gap which widens with increasing film thickness). This suggests that that polydomain FE competes with AFD modes. This effect is not as drastic as the competition between FE modes and monodomain [100] polarization however.
    
    \subsection{\label{morph:level2} Polarization morphologies}
    
\begin{figure*}
       \includegraphics[width=\linewidth]{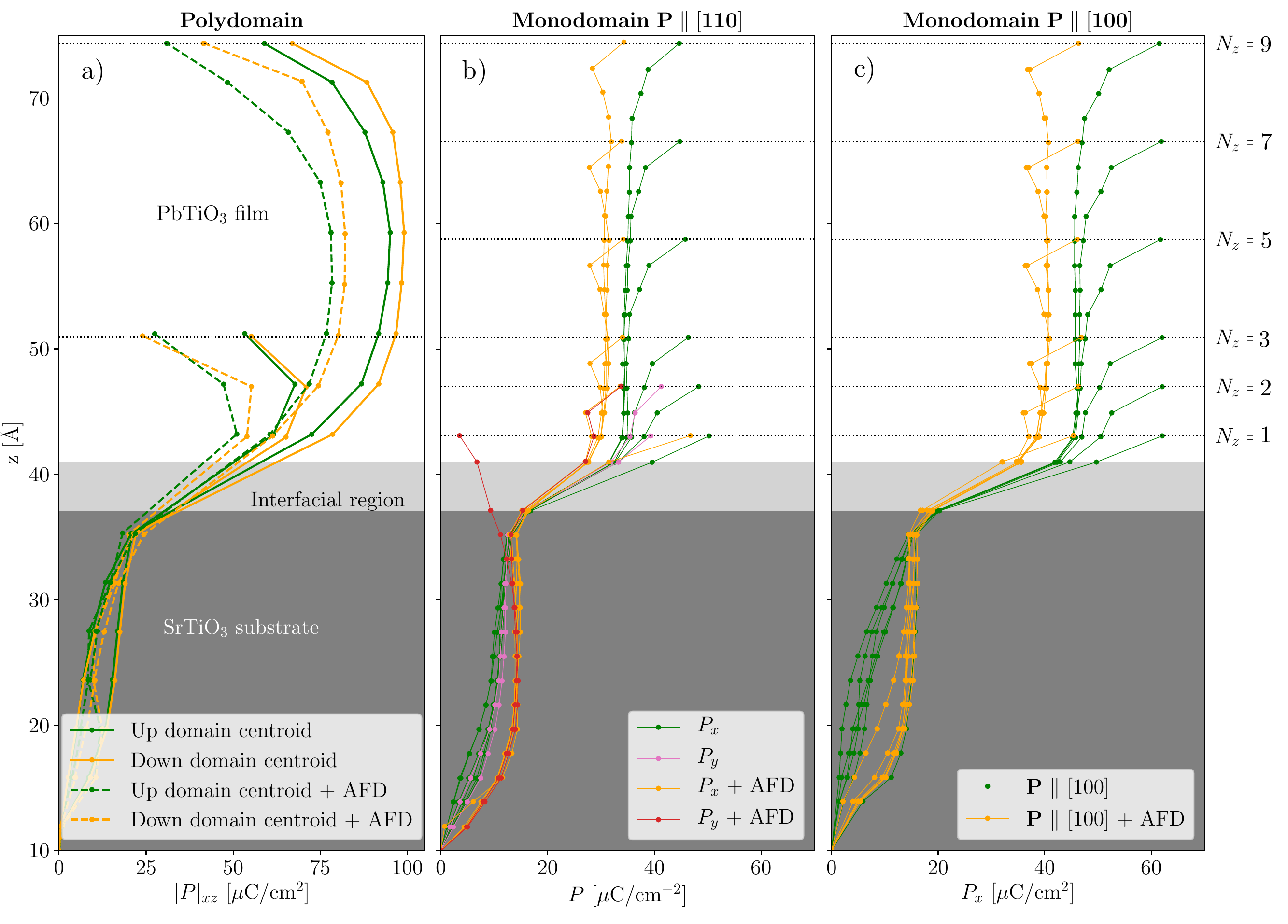}
        \caption{The local polarization profiles as a function of film vertical position $z$ for different film geometries with and without the interplay of AFD modes. a) The local polarization ($|P|_{xz} = \sqrt{P_x^2 + P_z^2}$, which is $\approx P_z$ at the domain centroid apart from the $N_z = 3$ film which has polar wave morphology) at the domain centroids for both the up and down domains of poly domain films. We display only $N_z = 3$ and $N_z = 9$ for clarity. b) The local polarization of films initially in the $\mathbf{P}$ $||$ [110] configuration. Both $P_x$ and $P_y$ components are shown for $N_z = 1$ and $N_z = 2$ (where there is a finite size effect). Otherwise, $P_x\approx P_y$. c) The local polarization ($P_x$) for films with $\mathbf{P}$ $||$ [100].}
        
        \label{fig:PolProf}
    \end{figure*}

\begin{figure}
       \includegraphics[width=\linewidth]{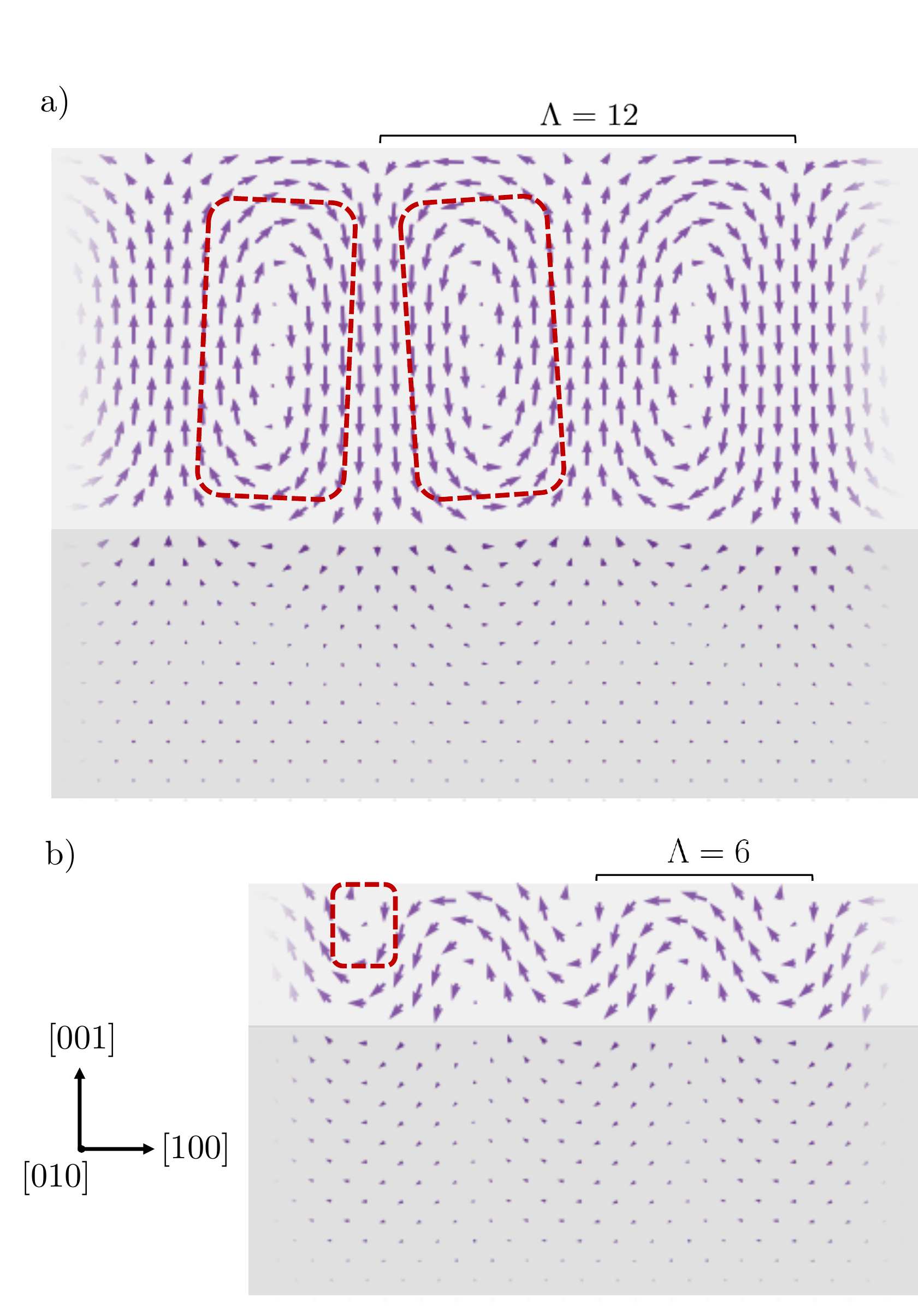}
        \caption{The local polarization vector fields in the x-z plane for two film thicknesses not including AFD modes. a) The flux-closure domains of the $N_z=9$, $\Lambda = 12$ film. The red area highlights a vortex/antivortex pair b) The polar wave morphology in the $N_z=3$, $\Lambda = 6$ film. The red area indicates a cylindrical chiral bubble.}
        
        \label{fig:VecField}
    \end{figure}

    In this section we analyse and compare the polar morphologies of the different films using a metric known as the \textit{local polarization}, $\mathbf{P}^{(i)}$. We define $\mathbf{P}^{(i)}$ per 5-atom perovskite unit cell using a linear relationship between the local mode and the Born effective charges $\mathbf{Z}^*$ in the manner first suggested by Resta \cite{Resta1993} (now used in similar works \cite{Meyer2002, AguadoPuente2012}). We discuss this method in more detail in the supporting information. A vector field of this quantity has been calculated for all polar structures presented in Figure \ref{fig:PolProf} and \ref{fig:VecField}.  
    
  In Figure \ref{fig:PolProf}a we show the Ti-centered local polarization along the up and down domain centroids. The domain centroid here is a string of Ti centered unit cells in the vertical direction located at the centre of a domain. It is at the down domain centroid that the maximal local polarization can be found, buried in the upper third of the PTO film. Indeed, there is a discrepancy in polarization between the up and down domains throughout the entire film, leading to a small net dipole moment in the [00$\bar{1}$] direction. This effect can be explained by the compositionally broken inversion symmetry present in even the highest symmetry films (the relaxed geometry described in Section \ref{para:level3}). The result is that $\mathbf{P} \parallel$ [00$\bar{1}$] is favored by a built-in bias field directed towards the substrate. While we find that this field is local to the first few PTO surface monolayers, we suggest that the enhanced local $\mathbf{P} \parallel$ [00$\bar{1}$] modes at the surface spread to the rest of the domain due to a finite correlation length associated with the polar atomic displacements \cite{Glinchuk2010}. This leads to an indentation of the substrate (as seen in figure \ref{fig:rumpling}) creating extra volume for the down domains and enhanced local tetragonality which mutually couples with the polarisation. The opposite argument is also true for the local $\mathbf{P} \parallel$ [001] modes within the up domain whereby the internal bias field is now depolarising. We predict that such a disparity between the up and down domains will diminish with increasing film thickness, tending to zero as the film thickness becomes much larger than the correlation length of the enhanced local polar modes. The scenario explained above is analogous to the observation of built-in bias fields present three component or \textit{tricolor} superlattices \cite{Sai2000, Warusawithana2003} where our third component is the vacuum region \par.
    
    We find that the maximal polarization increases with film thickness for the polydomain films. Although we only display $N_z = 3$ and $N_z = 9$ in Figure \ref{fig:PolProf}a, the rise is gradual when considering the maximal polarization of the $N_z = 5$ and $N_z = 7$ films also. This reflects the results of synchrotron X-ray diffraction \cite{Fong2004} whereby satellite peak intensity (indicative of domain polarization) is an increasing function of film thickness. Such a phenomenon is likely a result of decreasing depolarising field strength with increasing film thickness as is observed for thicker $\mathbf{P} \parallel$ [001] films \cite{Lichtensteiger2005}. It is notable that by $N_z = 9$, this maximal polarization (for the case \textit{without} AFD modes) exceeds the strained PTO bulk figure by 26\%. The same enhancement is not seen for the $N_z = 3$ which suffers a 12\% reduction. It can be seen for all film geometries that the influence of AFD modes tends to \textit{reduce} the polarization in the film. Figure \ref{fig:PolProf}a shows that allowing for AFD modes produces a local polarization reduction of  $\approx$ 15 $\mu C/cm^{2}$ for all films. This reduction, however, becomes more severe as we reach the surface where we suggest that the enhanced local rotational modes at the surface reconstruction compete strongly with the polarization. \par
    
    The majority of the relaxed polydomain structures, as seen in similar works, form the flux closure morphology (Figure \ref{fig:VecField}a). That is, the local polarization gradually rotates through 180$^{\circ}$ across the domain wall at the top of the film and rotates through -180$^{\circ}$ at the bottom as a mechanism for screening the depolarising field. The result is counter-tilting vortex-like domain walls (Figure \ref{fig:VecField}a, red areas), with a small vertical area for $\approx 0$ polarization at the vortex center. This domain morphology forms fully for the $N_z = 5, 7, 9$ films with and without the influence of AFD modes. We note that whilst these domain structures share similarities with other works, they do have some key differences. The vortex cores are not located centrally (in the z-direction) in the film. They are instead shifted towards the substrate. We also see that at the surface, flux is closed more abruptly than at the interface with the substrate. At the interface, the polarization penetrates ($\approx 2$ unit cells) into the substrate helping to close the flux. Together with the previously mentioned tendency for a stronger polarization in the [00$\bar{1}$] direction, this makes for a more asymmetrical morphology than those observed in the superlattice or free standing film arrangement \cite{Shimada2010, AguadoPuente2012}. In contrast to the work of Shimada \cite{Shimada2010}, the $N_z = 3$ film does not appear to form complete a flux closure domain structure. Examining Figure \ref{fig:VecField}b we see that the flux does not fully close at the interface with STO. Instead, the polar morphology is \textit{S}-like or wave-like, in this case orientated in the [$\bar{1}$00] direction giving the film a net in-plane macroscopic polarization. We deduce that at the surface, the flux \textit{does} close by analysing the displacements of the terminating PbO layer (whose polarization vectors are not calculable with our method as explained in the supporting information). This gives rise to small cylindrical vortices near the surface which are sometimes known as \textit{cylindrical chiral bubbles} (Figure \ref{fig:VecField}b, red area) \cite{Lu2018, Sichuga2011}. \par
    
    Such an instability has been predicted as an intermediate phase in phase field simulations under applied electric fields in the superlattice arrangement \cite{Damodaran2017} as well as at zero field in high-angle annular dark field (HAADF) images \cite{Lu2018}. This polar topology is also observed in thin PZT films as predicted with Monte-Carlo simulations of a first principles-based Hamiltonian \cite{Sichuga2011}. Although such a polar topology has not yet been observed in thin PTO films, we note that previous XRD studies do not explicitly rule out the coupling to the characteristic in-plane wavevectors for the $N_z=3$ film \cite{Stephenson2003, Streiffer2002, Fong2004}. In this work, the polar-wave morphology appears as an intermediate phase between full flux-closure domains and in-plane polarization like in reference \cite{Damodaran2017} but in the absence of an applied field. This is replaced by the built-in bias field emergent from the compositionally broken inversion symmetry. Since this bias field is directed to suppress up domains but enhance down domains, up domains now have an increased critical thicknesses for the absolute suppression of out-of-plane polarization compared to down domains. We suggest that at $N_z = 3$, we are below this critical thickness for up domains (where in-plane FE modes are now favored) but still above it in down domains. The resulting polar wave texture is then emergent from the closing of flux between $\mathbf{P} \parallel$ [00$\bar{1}$] and $\mathbf{P} \parallel$ [$\bar{1}$00] (although $\mathbf{P} \parallel$ [100] is equally favored) domains as shown in figure \ref{fig:VecField}b. This finding suggests that control over polar morphologies can be achieved in ultrathin films by careful engineering of the boundary conditions. Specifically, we suggest that the built-in bias field can be manipulated through the choice of substrate, film surface termination or use of overlayers. This principle of design offers a promising avenue for the manipulation of chiral order in low-dimensional devices operating through the control of toroidal moments. \par
    
    Figure \ref{fig:PolProf}b and \ref{fig:PolProf}c show the polarization profiles for the films initialised with uniform polarization oriented in the [110] and [100] directions respectively. For the former, we find that for most cases, $P_x \approx P_y$ (hence $\mathbf{P}$ remains aligned along [110]) apart from the thinnest $N_z = 1$ and 2 films. $N_z = 1$ in particular, when coupled with AFD modes, (red line) polarization rotates strongly to be mostly polarised in [001]. For both [110] and [100] polarised films, we see that these films become polar as we cross the STO/PTO interface, quickly adopting a bulk like value. Then, for films without AFD modes, we see a polar enhancement near the surface. For those with AFD modes, we see a polar reduction followed by enhancement near the surface contrasting with the strong reduction for the polydomain films. Computing the norm $\sqrt{P_x^2 + P_y^2}$ of the polarization for the [110] and [100] films in the bulk-like region shows us that the [110] films have a polarization slightly enhanced (+2$\mu C/cm^{2}$) compared to just [100] polarization. What is notable for the monodomain films is that (past a thickness of two unit cells at least) there is no trend for the behaviour of the local polarization for increasing film thickness. Each film shows the same bulk-like polarization then the same surface reduction or enhancement.

    \subsection{\label{surfrecon:level2}  The p($2 \times \Lambda$) surface reconstruction}
    
    \begin{figure}
       \includegraphics[width=\linewidth]{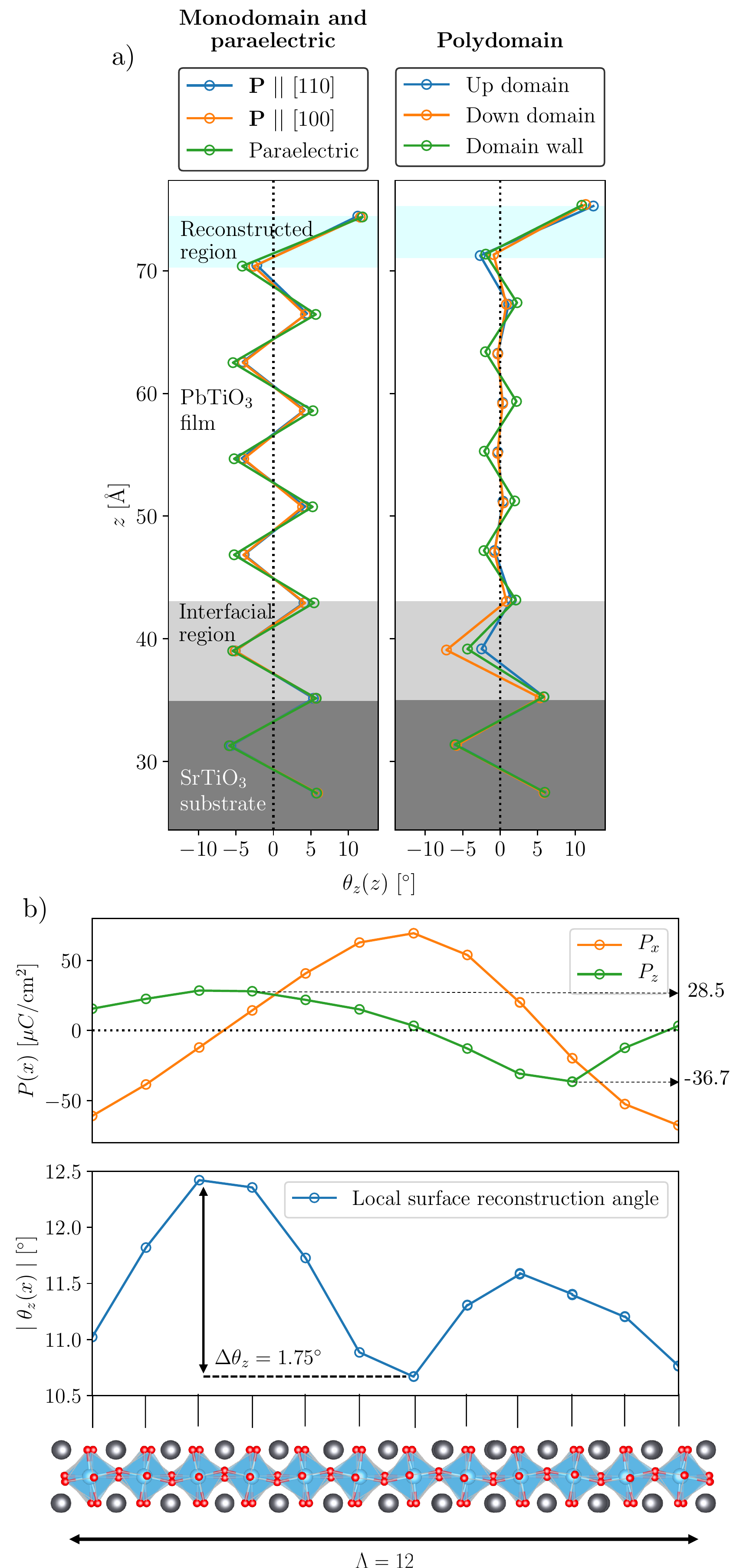}
        \caption{a) The $a^0a^0c^-$ TiO$_6$ octahedral rotation angle $\theta_z(z)$ for the different geometries of the $N_z = 9$ film. b) $\theta_z(x)$ and the $x$ and $z$ components of the local polarization $P(x)$ across a domain period $\Lambda$ of 12 unit cells for the $N_z = 9$ film. Since both the local polarization and rotations are centered on Ti sites, as a visual guide, an aerial view of the film looking down [00$\bar{1}$] is below a) aligned with the points on a)}. 
        \label{fig:SurfRecon}
    \end{figure}
    
    \begin{figure*}
       \includegraphics[width=\linewidth]{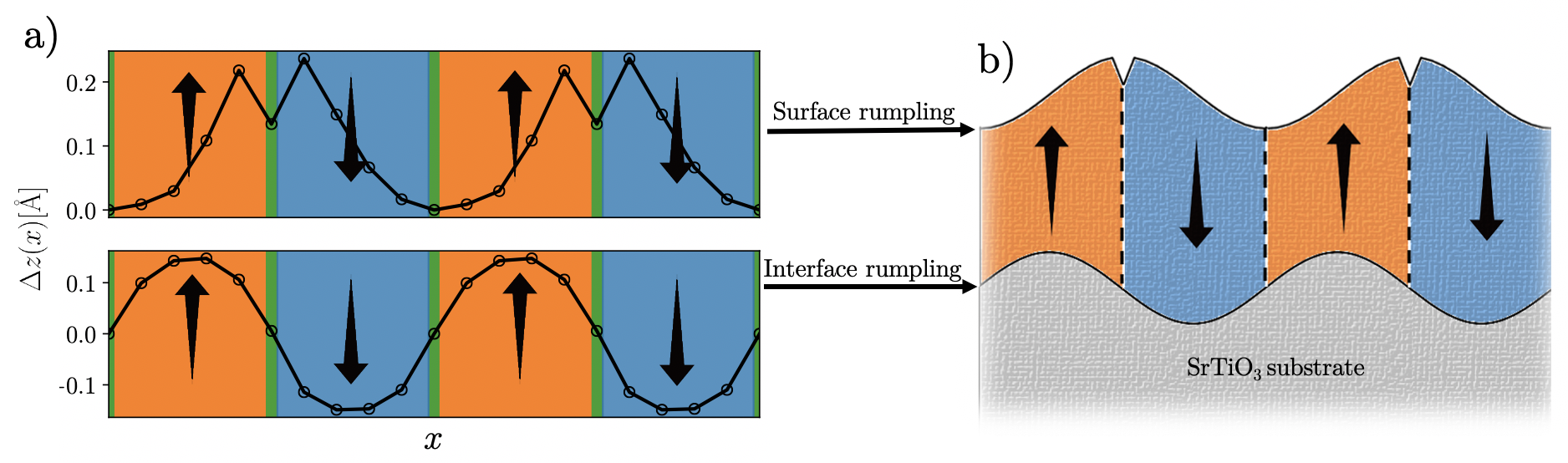}
        \caption{Surface and interface rumpling of the Pb cations for the $N_z = 7$ ($\Lambda = 10$) film \textit{without} AFD modes. a) The vertical deviation of the surface (upper) and interface (lower) Pb atoms from the Pb atom located at the first domain wall (along the x-axis). b) A schematic of the surface and interface rumpling including the boundary with the substrate.}
        \label{fig:rumpling}
    \end{figure*} 
    
    An analysis of the competition between local FE and AFD modes can provide valuable insight into the design of new low dimensional devices whereby FE and AFD modes can be tuned to enhance their functional properties. In this section, we analyse the interaction of FE and AFD modes with a focus on strong coupling within the reconstructed surface layers of the PbO terminated films. \par 
    
    Figure \ref{fig:SurfRecon} shows the local evolution of the $R_4^+$ octahedral rotation angle $\theta_z$ along the [001] direction (Figure \ref{fig:SurfRecon}a) and across a domain period $\Lambda$ in the [100] direction (Figure \ref{fig:SurfRecon}b) for the $N_z = 9$ ($\Lambda = 12$) film. The rotational behaviour in Figure \ref{fig:SurfRecon}a (left) is similar to the behaviour reported by Bungaro \cite{Bungaro2005} whereby the reconstructed area couples cooperatively with in-plane [100] polarization. We show that this also true for $\mathbf{P}$ $||$ [110] (polarization is locally enhanced at the surface as shown by Figure \ref{fig:PolProf}b and \ref{fig:PolProf}c) with the rotation pattern $\theta_z(z)$ being almost indistinguishable from the [100] polar film. We note that whilst this mutual AFD/FE enhancement is active at the surface, for the rest of the film, AFD/FE modes are mutually reduced when compared to the paraelectric structure and the bulk. The $\theta_z(z)$ trend for the polydomain films (Figure \ref{fig:SurfRecon}a, right) is more complex. Rotations are almost completely quenched (reduced to $\approx 2^{\circ}$) at the up and down domain centroids compared to the domain wall (apart from the reconstructed region). We suggest that this occurs since the maximal polarisation at the centroids outcompetes AFD modes. We see similar behaviour for all $N_z$ for the effects listed above, even for the strength of the reconstruction angle. This is remarkably resistant to finite size effects, even at a single PbO monolayer ($N_z = 0$), $\theta_z$ persists at around $12^{\circ}$ despite the change in chemical environment for the surface TiO$_6$ octahedra. \par
    
    Figure \ref{fig:SurfRecon}b shows a coupling between local polarization at the surface (upper) and the surface reconstruction angle (lower). Across a domain period $\Lambda$, we see that $\theta_z(x)$ modulates by $1.75^{\circ}$. $|\theta_z(x)|$ peaks close to $P_x = 0$ which is remarkable since this is precisely the \textit{opposite} behaviour to the monodomain in-plane films where FE and AFD mutually support one another. Since $P_x(x)$ and $P_z(x)$ are $\pi/2$ radians out of phase with each other in the surface layer, it can also be said that the peaks in $|\theta_z(x)|$ coincide with extrema in $|P_z(x)|$. We see also the the magnitude of $P_z(x)$ has an impact on the height of the peaks in $|\theta_z(x)|$. That is, since the down domains are more polar than the up domains, the larger down polarization in the surface layer reduces the $|\theta_z(x)|$ peak and vice versa for the for the smaller polarization in the up domain. We note that such AFD/FE couplings are different to those found in a recent shell model study of the free standing film \cite{Chapman2017}. In this study, the symmetrical [001] boundary conditions for the film modulate $\theta_z(x)$ over $\Lambda/2$ instead. We also note that whilst the strongest octahedral rotations $\theta_z$ are about the [001] axis, we also see smaller rotations of $\approx$ 2-3$^{\circ}$ about the [100] axis in the surface reconstruction. In addition to this, at the surface of the polar wave film for $N_z = 3$, we observe rotations about \textit{all three} pseudocubic axes in the surface reconstruction. The strongest rotations are still about the [001] axis however. \par
    
    The broken inversion symmetry leads to further structural effects for the polydomain films. We can measure the distortion of the films in the vertical direction by considering the Pb displacements in the terminating PbO layer with the vaccuum and at the STO interface. \ref{fig:rumpling} shows an exemplar calculation for the $N_z = 7$ ($\Lambda=10$) film. The other film thicknesses all resemble the behaviour of this film. With the exception of a small dip in $\Delta z$ at the domain wall for the surface rumpling (a mechanism to reduce the domain wall formation energy), we see that interface and surface rumpling behave spatially like two sinusoids $\pi/2$ radians out-of-phase with one-another, reflected about the [100] direction. This is in contrast to the PTO/STO superlattice configuration \cite{AguadoPuente2012} where the intact inversion symmetry preserves symmetrical rumpling (in-phase sinusoids). Due to the periodicity of octahedral rotations and Pb cation surface rumpling in the [001] direction, it is clear that for the polydomain films we must reconsider the labelling of the surface reconstruction. For monodomain and paraelectric films, the Wood's notation \cite{prutton1994introduction} of $c(2 \times 2)$ remains correct but an analysis of the symmetry for the the polydomain case reveals this must alter to become $p(2 \times \Lambda)$. We suggest that these fine FE/AFD interactions could now be observable in experiment thanks to recent advances in integrated differential phase contrast (iDPC) imaging \cite{Sun2019}. In contrast to HAADF, iDPC images yield the positions of the metal cations \textit{and} oxygen anions resolved at a subunit cell level \cite{Sun2019}. This allows for the direct measurement of local FE and AFD modes offering an exciting new avenue for direct comparison with atomistic results.

    \section{\label{summary:level1}  Conclusions}
    
    We have used large-scale DFT calculations to simulate the energetics, interaction with AFD modes and other structural features for various polar morphologies present in thin PTO films on STO substrates. Our method is successful in providing accurate first principles results for systems comprised of a few thousand atoms, well beyond what is feasible with traditional plane-wave based methods, venturing towards systems sizes usually simulated with Monte Carlo or phase field techniques. This method has allowed for the explicit simulation of the STO substrate, multiple ferroelectric domains and doubled periodicity in the [010] direction used include AFD modes and the surface reconstruction. This has ensured that our simulations represent realistic experimental conditions when compared to previous works which neglect one or more of these features. Such simulations can be performed on standard HPC systems with high accuracy calculations of $\approx 1000-2000$ atoms becoming feasible using $\approx 200-500$ physical cores.\par
    
    We have demonstrated the stability of the polydomain film geometry compared to monodomain phases. We have found that the polydomain case becomes more energetically favorable between 4-5 unit cells in thickness, close to the experimental observation at 3 unit cells. We find that the general effect of including AFD modes is to lower the energy significantly (in most cases more than the FE distortion) and to suppress the amplitudes of local polar modes (apart from at the surface of monodomain in-plane FE films, where they are mutually cooperative). \par
    
    We show that polydomain films display the flux-closure domain morphology for $N_z = 5, 7$ and 9 whilst the $N_z = 3$ film shows the similar polar wave morphology with cylindrical chiral bubbles as an intermediate phase between full flux closure domains and in-plane ferroelectricity. We find that local polarization is enhanced at the domain centroids when compared to PTO bulk; a trend which increases as a function of $N_z$. Most notably, down domains feature enhanced local polar modes promoted by the internal bias field born of the compositionally broken inversion symmetry. Equally, this bias field acts to depolarise up domains leading to different critical thicknesses for the total suppression of out-of-plane ferroelectricity for the two domains. Since these critical thicknesses are the outcome of the strength and direction of the bias field, engineering this with a careful choice of substrate, surface termination or overlayers allows for control over polar textures at the nanoscale. This finding is especially important for next-generation functional devices reliant upon the control of toroidal order.  \par
    
    There are consequences also for the periodicity of the AFD surface reconstruction. While we find that the reconstruction for the monodomain in-plane FE and paraelectric films is $c(2 \times 2)$, we find that coupling between surface polarization and AFD modes in the polydomain geometry modifies this. In addition to surface rumpling of Pb cations, the surface reconstruction angle modulates up to 1.75$^{\circ}$ across a domain period $\Lambda$. We then suggest that for the polydomain films, the correct label for the AFD surface reconstruction is $p(2 \times \Lambda)$ in Wood's notation. This provides direct evidence that the strength of AFD modes can be locally controlled by the strength of FE modes and vice versa. Such knowledge could motivate new principles of design in low dimensional functional devices whereby FE and AFD modes are locally tuned by their interactions. \par
    
    We suggest that the MSSF method implemented within the \textsc{CONQUEST} code can now be used to solve a plethora of problems within the perovskite oxides. This could include the simulating other potentially possible polar morphologies in the PTO/STO system such as skyrmion phases \cite{Hong2018} and disclinations \cite{Damodaran2017}. Since our method is general, we can extend to other problems in the perovskite oxides (and beyond) such as realistic defect concentrations and highly disordered configurations the popular solid solution families AB$_{x}$C$_{1-x}$O$_3$, $(1-x)$ABO$_3-x$CDO$_3$ where DFT methods used to circumvent the need for large supercell calculations fail in the reproduction of local structural distortions \cite{Baker2019}.

\section*{Acknowledgements}

We are grateful for computational support from the UK
Materials and Molecular Modelling Hub, which is partially
funded by EPSRC (EP/P020194), for which access was obtained via the UKCP consortium and funded by EPSRC Grant
Ref. No. EP/P022561/1. This work also used the ARCHER
UK National Supercomputing Service funded by the UKCP
consortium EPSRC Grant Ref. No. EP/P022561/1. We are also thankful for the oversight of Pavlo Zubko and Marios Hadjimichael in reviewing this manuscript and thank Pablo Aguado-Puente for his guidance in the preparation of local polarization vector fields.

\bibliography{PTOFilm}

\begin{thebibliography}{77}%
\makeatletter
\providecommand \@ifxundefined [1]{%
 \@ifx{#1\undefined}
}%
\providecommand \@ifnum [1]{%
 \ifnum #1\expandafter \@firstoftwo
 \else \expandafter \@secondoftwo
 \fi
}%
\providecommand \@ifx [1]{%
 \ifx #1\expandafter \@firstoftwo
 \else \expandafter \@secondoftwo
 \fi
}%
\providecommand \natexlab [1]{#1}%
\providecommand \enquote  [1]{``#1''}%
\providecommand \bibnamefont  [1]{#1}%
\providecommand \bibfnamefont [1]{#1}%
\providecommand \citenamefont [1]{#1}%
\providecommand \href@noop [0]{\@secondoftwo}%
\providecommand \href [0]{\begingroup \@sanitize@url \@href}%
\providecommand \@href[1]{\@@startlink{#1}\@@href}%
\providecommand \@@href[1]{\endgroup#1\@@endlink}%
\providecommand \@sanitize@url [0]{\catcode `\\12\catcode `\$12\catcode
  `\&12\catcode `\#12\catcode `\^12\catcode `\_12\catcode `\%12\relax}%
\providecommand \@@startlink[1]{}%
\providecommand \@@endlink[0]{}%
\providecommand \url  [0]{\begingroup\@sanitize@url \@url }%
\providecommand \@url [1]{\endgroup\@href {#1}{\urlprefix }}%
\providecommand \urlprefix  [0]{URL }%
\providecommand \Eprint [0]{\href }%
\providecommand \doibase [0]{http://dx.doi.org/}%
\providecommand \selectlanguage [0]{\@gobble}%
\providecommand \bibinfo  [0]{\@secondoftwo}%
\providecommand \bibfield  [0]{\@secondoftwo}%
\providecommand \translation [1]{[#1]}%
\providecommand \BibitemOpen [0]{}%
\providecommand \bibitemStop [0]{}%
\providecommand \bibitemNoStop [0]{.\EOS\space}%
\providecommand \EOS [0]{\spacefactor3000\relax}%
\providecommand \BibitemShut  [1]{\csname bibitem#1\endcsname}%
\let\auto@bib@innerbib\@empty
\bibitem [{\citenamefont {Schlom}\ \emph {et~al.}(2008)\citenamefont {Schlom},
  \citenamefont {Chen}, \citenamefont {Pan}, \citenamefont {Schmehl},\ and\
  \citenamefont {Zurbuchen}}]{Schlom2008}%
  \BibitemOpen
  \bibfield  {author} {\bibinfo {author} {\bibfnamefont {D.~G.}\ \bibnamefont
  {Schlom}}, \bibinfo {author} {\bibfnamefont {L.-Q.}\ \bibnamefont {Chen}},
  \bibinfo {author} {\bibfnamefont {X.}~\bibnamefont {Pan}}, \bibinfo {author}
  {\bibfnamefont {A.}~\bibnamefont {Schmehl}}, \ and\ \bibinfo {author}
  {\bibfnamefont {M.~A.}\ \bibnamefont {Zurbuchen}},\ }\href
  {https://doi.org/10.1111/j.1551-2916.2008.02556.x} {\bibfield  {journal}
  {\bibinfo  {journal} {J. Am. Ceram. Soc.}\ }\textbf {\bibinfo {volume}
  {91}},\ \bibinfo {pages} {2429} (\bibinfo {year} {2008})}\BibitemShut
  {NoStop}%
\bibitem [{\citenamefont {Willmott}(2004)}]{Willmott2004}%
  \BibitemOpen
  \bibfield  {author} {\bibinfo {author} {\bibfnamefont {P.}~\bibnamefont
  {Willmott}},\ }\href {https://doi.org/10.1016/j.progsurf.2004.06.001}
  {\bibfield  {journal} {\bibinfo  {journal} {Prog. Surf. Sci.}\ }\textbf
  {\bibinfo {volume} {76}},\ \bibinfo {pages} {163} (\bibinfo {year}
  {2004})}\BibitemShut {NoStop}%
\bibitem [{\citenamefont {Kobayashi}\ \emph {et~al.}(1998)\citenamefont
  {Kobayashi}, \citenamefont {Kimura}, \citenamefont {Sawada}, \citenamefont
  {Terakura},\ and\ \citenamefont {Tokura}}]{Kobayashi1998}%
  \BibitemOpen
  \bibfield  {author} {\bibinfo {author} {\bibfnamefont {K.-I.}\ \bibnamefont
  {Kobayashi}}, \bibinfo {author} {\bibfnamefont {T.}~\bibnamefont {Kimura}},
  \bibinfo {author} {\bibfnamefont {H.}~\bibnamefont {Sawada}}, \bibinfo
  {author} {\bibfnamefont {K.}~\bibnamefont {Terakura}}, \ and\ \bibinfo
  {author} {\bibfnamefont {Y.}~\bibnamefont {Tokura}},\ }\href
  {https://doi.org/10.1038/27167} {\bibfield  {journal} {\bibinfo  {journal}
  {Nature}\ }\textbf {\bibinfo {volume} {395}},\ \bibinfo {pages} {677}
  (\bibinfo {year} {1998})}\BibitemShut {NoStop}%
\bibitem [{\citenamefont {Sleight}\ \emph {et~al.}(1993)\citenamefont
  {Sleight}, \citenamefont {Gillson},\ and\ \citenamefont
  {Bierstedt}}]{Sleight1993}%
  \BibitemOpen
  \bibfield  {author} {\bibinfo {author} {\bibfnamefont {A.~W.}\ \bibnamefont
  {Sleight}}, \bibinfo {author} {\bibfnamefont {J.~L.}\ \bibnamefont
  {Gillson}}, \ and\ \bibinfo {author} {\bibfnamefont {P.~E.}\ \bibnamefont
  {Bierstedt}},\ }in\ \href {https://doi.org/10.1007/978-94-011-1622-0_30}
  {\emph {\bibinfo {booktitle} {Ten Years of Superconductivity:
  1980{\textendash}1990}}}\ (\bibinfo  {publisher} {Springer Netherlands},\
  \bibinfo {year} {1993})\ pp.\ \bibinfo {pages} {257--258}\BibitemShut
  {NoStop}%
\bibitem [{\citenamefont {Hwang}\ \emph {et~al.}(2012)\citenamefont {Hwang},
  \citenamefont {Iwasa}, \citenamefont {Kawasaki}, \citenamefont {Keimer},
  \citenamefont {Nagaosa},\ and\ \citenamefont {Tokura}}]{Hwang2012}%
  \BibitemOpen
  \bibfield  {author} {\bibinfo {author} {\bibfnamefont {H.~Y.}\ \bibnamefont
  {Hwang}}, \bibinfo {author} {\bibfnamefont {Y.}~\bibnamefont {Iwasa}},
  \bibinfo {author} {\bibfnamefont {M.}~\bibnamefont {Kawasaki}}, \bibinfo
  {author} {\bibfnamefont {B.}~\bibnamefont {Keimer}}, \bibinfo {author}
  {\bibfnamefont {N.}~\bibnamefont {Nagaosa}}, \ and\ \bibinfo {author}
  {\bibfnamefont {Y.}~\bibnamefont {Tokura}},\ }\href
  {https://doi.org/10.1038/nmat3223} {\bibfield  {journal} {\bibinfo  {journal}
  {Nature Mat.}\ }\textbf {\bibinfo {volume} {11}},\ \bibinfo {pages} {103}
  (\bibinfo {year} {2012})}\BibitemShut {NoStop}%
\bibitem [{\citenamefont {Yin}\ \emph {et~al.}(2015)\citenamefont {Yin},
  \citenamefont {Aguado-Puente}, \citenamefont {Qu},\ and\ \citenamefont
  {Artacho}}]{Yin2015}%
  \BibitemOpen
  \bibfield  {author} {\bibinfo {author} {\bibfnamefont {B.}~\bibnamefont
  {Yin}}, \bibinfo {author} {\bibfnamefont {P.}~\bibnamefont {Aguado-Puente}},
  \bibinfo {author} {\bibfnamefont {S.}~\bibnamefont {Qu}}, \ and\ \bibinfo
  {author} {\bibfnamefont {E.}~\bibnamefont {Artacho}},\ }\href
  {https://doi.org/10.1103/physrevb.92.115406} {\bibfield  {journal} {\bibinfo
  {journal} {Phys. Rev. B}\ }\textbf {\bibinfo {volume} {92}},\ \bibinfo
  {pages} {115406} (\bibinfo {year} {2015})}\BibitemShut {NoStop}%
\bibitem [{\citenamefont {Zubko}\ \emph {et~al.}(2016)\citenamefont {Zubko},
  \citenamefont {Wojde{\l}}, \citenamefont {Hadjimichael}, \citenamefont
  {Fernandez-Pena}, \citenamefont {Sen{\'{e}}}, \citenamefont {Luk'yanchuk},
  \citenamefont {Triscone},\ and\ \citenamefont
  {{\'{I}}{\~{n}}iguez}}]{zubko2016negative}%
  \BibitemOpen
  \bibfield  {author} {\bibinfo {author} {\bibfnamefont {P.}~\bibnamefont
  {Zubko}}, \bibinfo {author} {\bibfnamefont {J.~C.}\ \bibnamefont
  {Wojde{\l}}}, \bibinfo {author} {\bibfnamefont {M.}~\bibnamefont
  {Hadjimichael}}, \bibinfo {author} {\bibfnamefont {S.}~\bibnamefont
  {Fernandez-Pena}}, \bibinfo {author} {\bibfnamefont {A.}~\bibnamefont
  {Sen{\'{e}}}}, \bibinfo {author} {\bibfnamefont {I.}~\bibnamefont
  {Luk'yanchuk}}, \bibinfo {author} {\bibfnamefont {J.-M.}\ \bibnamefont
  {Triscone}}, \ and\ \bibinfo {author} {\bibfnamefont {J.}~\bibnamefont
  {{\'{I}}{\~{n}}iguez}},\ }\href {https://doi.org/10.1038/nature17659}
  {\bibfield  {journal} {\bibinfo  {journal} {Nature}\ }\textbf {\bibinfo
  {volume} {534}},\ \bibinfo {pages} {524} (\bibinfo {year}
  {2016})}\BibitemShut {NoStop}%
\bibitem [{\citenamefont {Dawber}\ \emph {et~al.}(2005)\citenamefont {Dawber},
  \citenamefont {Rabe},\ and\ \citenamefont {Scott}}]{Dawber2005}%
  \BibitemOpen
  \bibfield  {author} {\bibinfo {author} {\bibfnamefont {M.}~\bibnamefont
  {Dawber}}, \bibinfo {author} {\bibfnamefont {K.~M.}\ \bibnamefont {Rabe}}, \
  and\ \bibinfo {author} {\bibfnamefont {J.~F.}\ \bibnamefont {Scott}},\ }\href
  {https://doi.org/10.1103/revmodphys.77.1083} {\bibfield  {journal} {\bibinfo
  {journal} {Rev. Mod. Phys.}\ }\textbf {\bibinfo {volume} {77}},\ \bibinfo
  {pages} {1083} (\bibinfo {year} {2005})}\BibitemShut {NoStop}%
\bibitem [{\citenamefont {Mannhart}\ and\ \citenamefont
  {Schlom}(2010)}]{Mannhart2010}%
  \BibitemOpen
  \bibfield  {author} {\bibinfo {author} {\bibfnamefont {J.}~\bibnamefont
  {Mannhart}}\ and\ \bibinfo {author} {\bibfnamefont {D.~G.}\ \bibnamefont
  {Schlom}},\ }\href {https://doi.org/10.1126/science.1181862} {\bibfield
  {journal} {\bibinfo  {journal} {Science}\ }\textbf {\bibinfo {volume}
  {327}},\ \bibinfo {pages} {1607} (\bibinfo {year} {2010})}\BibitemShut
  {NoStop}%
\bibitem [{\citenamefont {Zubko}\ \emph {et~al.}(2011)\citenamefont {Zubko},
  \citenamefont {Gariglio}, \citenamefont {Gabay}, \citenamefont {Ghosez},\
  and\ \citenamefont {Triscone}}]{Zubko2011}%
  \BibitemOpen
  \bibfield  {author} {\bibinfo {author} {\bibfnamefont {P.}~\bibnamefont
  {Zubko}}, \bibinfo {author} {\bibfnamefont {S.}~\bibnamefont {Gariglio}},
  \bibinfo {author} {\bibfnamefont {M.}~\bibnamefont {Gabay}}, \bibinfo
  {author} {\bibfnamefont {P.}~\bibnamefont {Ghosez}}, \ and\ \bibinfo {author}
  {\bibfnamefont {J.-M.}\ \bibnamefont {Triscone}},\ }\href
  {https://doi.org/10.1146/annurev-conmatphys-062910-140445} {\bibfield
  {journal} {\bibinfo  {journal} {Annu. Rev. Condens. Matter Phys.}\ }\textbf
  {\bibinfo {volume} {2}},\ \bibinfo {pages} {141} (\bibinfo {year}
  {2011})}\BibitemShut {NoStop}%
\bibitem [{\citenamefont {Lu}\ \emph {et~al.}(2018)\citenamefont {Lu},
  \citenamefont {Nahas}, \citenamefont {Liu}, \citenamefont {Du}, \citenamefont
  {Jiang}, \citenamefont {Ren}, \citenamefont {Wang}, \citenamefont {Jin},
  \citenamefont {Prokhorenko}, \citenamefont {Jia},\ and\ \citenamefont
  {Bellaiche}}]{Lu2018}%
  \BibitemOpen
  \bibfield  {author} {\bibinfo {author} {\bibfnamefont {L.}~\bibnamefont
  {Lu}}, \bibinfo {author} {\bibfnamefont {Y.}~\bibnamefont {Nahas}}, \bibinfo
  {author} {\bibfnamefont {M.}~\bibnamefont {Liu}}, \bibinfo {author}
  {\bibfnamefont {H.}~\bibnamefont {Du}}, \bibinfo {author} {\bibfnamefont
  {Z.}~\bibnamefont {Jiang}}, \bibinfo {author} {\bibfnamefont
  {S.}~\bibnamefont {Ren}}, \bibinfo {author} {\bibfnamefont {D.}~\bibnamefont
  {Wang}}, \bibinfo {author} {\bibfnamefont {L.}~\bibnamefont {Jin}}, \bibinfo
  {author} {\bibfnamefont {S.}~\bibnamefont {Prokhorenko}}, \bibinfo {author}
  {\bibfnamefont {C.-L.}\ \bibnamefont {Jia}}, \ and\ \bibinfo {author}
  {\bibfnamefont {L.}~\bibnamefont {Bellaiche}},\ }\href {\doibase
  10.1103/physrevlett.120.177601} {\bibfield  {journal} {\bibinfo  {journal}
  {Phys. Rev. Lett.}\ }\textbf {\bibinfo {volume} {120}},\ \bibinfo {pages}
  {177601} (\bibinfo {year} {2018})}\BibitemShut {NoStop}%
\bibitem [{\citenamefont {Sichuga}\ and\ \citenamefont
  {Bellaiche}(2011)}]{Sichuga2011}%
  \BibitemOpen
  \bibfield  {author} {\bibinfo {author} {\bibfnamefont {D.}~\bibnamefont
  {Sichuga}}\ and\ \bibinfo {author} {\bibfnamefont {L.}~\bibnamefont
  {Bellaiche}},\ }\href {\doibase 10.1103/physrevlett.106.196102} {\bibfield
  {journal} {\bibinfo  {journal} {Phys. Rev. Lett.}\ }\textbf {\bibinfo
  {volume} {106}},\ \bibinfo {pages} {196102} (\bibinfo {year}
  {2011})}\BibitemShut {NoStop}%
\bibitem [{\citenamefont {Damodaran}\ \emph {et~al.}(2017)\citenamefont
  {Damodaran}, \citenamefont {Clarkson}, \citenamefont {Hong}, \citenamefont
  {Liu}, \citenamefont {Yadav}, \citenamefont {Nelson}, \citenamefont {Hsu},
  \citenamefont {McCarter}, \citenamefont {Park}, \citenamefont {Kravtsov},
  \citenamefont {Farhan} \emph {et~al.}}]{Damodaran2017}%
  \BibitemOpen
  \bibfield  {author} {\bibinfo {author} {\bibfnamefont {A.~R.}\ \bibnamefont
  {Damodaran}}, \bibinfo {author} {\bibfnamefont {J.~D.}\ \bibnamefont
  {Clarkson}}, \bibinfo {author} {\bibfnamefont {Z.}~\bibnamefont {Hong}},
  \bibinfo {author} {\bibfnamefont {H.}~\bibnamefont {Liu}}, \bibinfo {author}
  {\bibfnamefont {A.~K.}\ \bibnamefont {Yadav}}, \bibinfo {author}
  {\bibfnamefont {C.~T.}\ \bibnamefont {Nelson}}, \bibinfo {author}
  {\bibfnamefont {S.-L.}\ \bibnamefont {Hsu}}, \bibinfo {author} {\bibfnamefont
  {M.~R.}\ \bibnamefont {McCarter}}, \bibinfo {author} {\bibfnamefont {K.-D.}\
  \bibnamefont {Park}}, \bibinfo {author} {\bibfnamefont {V.}~\bibnamefont
  {Kravtsov}}, \bibinfo {author} {\bibfnamefont {A.}~\bibnamefont {Farhan}},
  \emph {et~al.},\ }\href {\doibase 10.1038/nmat4951} {\bibfield  {journal}
  {\bibinfo  {journal} {Nat. Mater.}\ }\textbf {\bibinfo {volume} {16}},\
  \bibinfo {pages} {1003} (\bibinfo {year} {2017})}\BibitemShut {NoStop}%
\bibitem [{\citenamefont {Hong}\ and\ \citenamefont {Chen}(2018)}]{Hong2018}%
  \BibitemOpen
  \bibfield  {author} {\bibinfo {author} {\bibfnamefont {Z.}~\bibnamefont
  {Hong}}\ and\ \bibinfo {author} {\bibfnamefont {L.-Q.}\ \bibnamefont
  {Chen}},\ }\href {\doibase 10.1016/j.actamat.2018.04.022} {\bibfield
  {journal} {\bibinfo  {journal} {Acta Mater.}\ }\textbf {\bibinfo {volume}
  {152}},\ \bibinfo {pages} {155} (\bibinfo {year} {2018})}\BibitemShut
  {NoStop}%
\bibitem [{\citenamefont {Du}\ \emph {et~al.}(2019)\citenamefont {Du},
  \citenamefont {Zhang}, \citenamefont {Dai}, \citenamefont {Zhou},
  \citenamefont {Xie}, \citenamefont {Ren}, \citenamefont {Tian}, \citenamefont
  {Chen}, \citenamefont {Tendeloo},\ and\ \citenamefont {Zhang}}]{Du2019}%
  \BibitemOpen
  \bibfield  {author} {\bibinfo {author} {\bibfnamefont {K.}~\bibnamefont
  {Du}}, \bibinfo {author} {\bibfnamefont {M.}~\bibnamefont {Zhang}}, \bibinfo
  {author} {\bibfnamefont {C.}~\bibnamefont {Dai}}, \bibinfo {author}
  {\bibfnamefont {Z.~N.}\ \bibnamefont {Zhou}}, \bibinfo {author}
  {\bibfnamefont {Y.~W.}\ \bibnamefont {Xie}}, \bibinfo {author} {\bibfnamefont
  {Z.~H.}\ \bibnamefont {Ren}}, \bibinfo {author} {\bibfnamefont
  {H.}~\bibnamefont {Tian}}, \bibinfo {author} {\bibfnamefont {L.~Q.}\
  \bibnamefont {Chen}}, \bibinfo {author} {\bibfnamefont {G.~V.}\ \bibnamefont
  {Tendeloo}}, \ and\ \bibinfo {author} {\bibfnamefont {Z.}~\bibnamefont
  {Zhang}},\ }\href {\doibase 10.1038/s41467-019-12864-5} {\bibfield  {journal}
  {\bibinfo  {journal} {Nat. Commun.}\ }\textbf {\bibinfo {volume} {10}},\
  \bibinfo {pages} {4864} (\bibinfo {year} {2019})}\BibitemShut {NoStop}%
\bibitem [{\citenamefont {Prosandeev}\ \emph {et~al.}(2008)\citenamefont
  {Prosandeev}, \citenamefont {Ponomareva}, \citenamefont {Naumov},
  \citenamefont {Kornev},\ and\ \citenamefont {Bellaiche}}]{Prosandeev2008}%
  \BibitemOpen
  \bibfield  {author} {\bibinfo {author} {\bibfnamefont {S.}~\bibnamefont
  {Prosandeev}}, \bibinfo {author} {\bibfnamefont {I.}~\bibnamefont
  {Ponomareva}}, \bibinfo {author} {\bibfnamefont {I.}~\bibnamefont {Naumov}},
  \bibinfo {author} {\bibfnamefont {I.}~\bibnamefont {Kornev}}, \ and\ \bibinfo
  {author} {\bibfnamefont {L.}~\bibnamefont {Bellaiche}},\ }\href {\doibase
  10.1088/0953-8984/20/19/193201} {\bibfield  {journal} {\bibinfo  {journal}
  {J. Phys. Condens. Matter}\ }\textbf {\bibinfo {volume} {20}},\ \bibinfo
  {pages} {193201} (\bibinfo {year} {2008})}\BibitemShut {NoStop}%
\bibitem [{\citenamefont {Chen}\ \emph {et~al.}(2015)\citenamefont {Chen},
  \citenamefont {Zheng},\ and\ \citenamefont {Wang}}]{Chen2015}%
  \BibitemOpen
  \bibfield  {author} {\bibinfo {author} {\bibfnamefont {W.~J.}\ \bibnamefont
  {Chen}}, \bibinfo {author} {\bibfnamefont {Y.}~\bibnamefont {Zheng}}, \ and\
  \bibinfo {author} {\bibfnamefont {B.}~\bibnamefont {Wang}},\ }\href {\doibase
  10.1038/srep11165} {\bibfield  {journal} {\bibinfo  {journal} {Sci. Rep.}\
  }\textbf {\bibinfo {volume} {5}},\ \bibinfo {pages} {11165} (\bibinfo {year}
  {2015})}\BibitemShut {NoStop}%
\bibitem [{\citenamefont {Nakata}\ \emph {et~al.}(2020)\citenamefont {Nakata},
  \citenamefont {Baker}, \citenamefont {Mujahed}, \citenamefont {Poulton},
  \citenamefont {Arapan}, \citenamefont {Lin}, \citenamefont {Raza},
  \citenamefont {Yadav}, \citenamefont {Truflandier}, \citenamefont
  {Miyazaki},\ and\ \citenamefont {Bowler}}]{Nakata2020large}%
  \BibitemOpen
  \bibfield  {author} {\bibinfo {author} {\bibfnamefont {A.}~\bibnamefont
  {Nakata}}, \bibinfo {author} {\bibfnamefont {J.~S.}\ \bibnamefont {Baker}},
  \bibinfo {author} {\bibfnamefont {S.~Y.}\ \bibnamefont {Mujahed}}, \bibinfo
  {author} {\bibfnamefont {J.~T.~L.}\ \bibnamefont {Poulton}}, \bibinfo
  {author} {\bibfnamefont {S.}~\bibnamefont {Arapan}}, \bibinfo {author}
  {\bibfnamefont {J.}~\bibnamefont {Lin}}, \bibinfo {author} {\bibfnamefont
  {Z.}~\bibnamefont {Raza}}, \bibinfo {author} {\bibfnamefont {S.}~\bibnamefont
  {Yadav}}, \bibinfo {author} {\bibfnamefont {L.}~\bibnamefont {Truflandier}},
  \bibinfo {author} {\bibfnamefont {T.}~\bibnamefont {Miyazaki}}, \ and\
  \bibinfo {author} {\bibfnamefont {D.~R.}\ \bibnamefont {Bowler}},\ }\href
  {\doibase 10.1063/5.0005074} {\bibfield  {journal} {\bibinfo  {journal} {J.
  Chem. Phys.}\ }\textbf {\bibinfo {volume} {152}},\ \bibinfo {pages} {164112}
  (\bibinfo {year} {2020})}\BibitemShut {NoStop}%
\bibitem [{\citenamefont {Nakata}\ \emph {et~al.}(2014)\citenamefont {Nakata},
  \citenamefont {Bowler},\ and\ \citenamefont {Miyazaki}}]{Nakata2014}%
  \BibitemOpen
  \bibfield  {author} {\bibinfo {author} {\bibfnamefont {A.}~\bibnamefont
  {Nakata}}, \bibinfo {author} {\bibfnamefont {D.~R.}\ \bibnamefont {Bowler}},
  \ and\ \bibinfo {author} {\bibfnamefont {T.}~\bibnamefont {Miyazaki}},\
  }\href {https://doi.org/10.1021/ct5004934} {\bibfield  {journal} {\bibinfo
  {journal} {J. Chem. Theory Comput.}\ }\textbf {\bibinfo {volume} {10}},\
  \bibinfo {pages} {4813} (\bibinfo {year} {2014})}\BibitemShut {NoStop}%
\bibitem [{\citenamefont {Gu}\ \emph {et~al.}(2018)\citenamefont {Gu},
  \citenamefont {Scarbrough}, \citenamefont {Yang}, \citenamefont
  {{\'{I}}{\~{n}}iguez}, \citenamefont {Bellaiche},\ and\ \citenamefont
  {Xiang}}]{Gu2018}%
  \BibitemOpen
  \bibfield  {author} {\bibinfo {author} {\bibfnamefont {T.}~\bibnamefont
  {Gu}}, \bibinfo {author} {\bibfnamefont {T.}~\bibnamefont {Scarbrough}},
  \bibinfo {author} {\bibfnamefont {Y.}~\bibnamefont {Yang}}, \bibinfo {author}
  {\bibfnamefont {J.}~\bibnamefont {{\'{I}}{\~{n}}iguez}}, \bibinfo {author}
  {\bibfnamefont {L.}~\bibnamefont {Bellaiche}}, \ and\ \bibinfo {author}
  {\bibfnamefont {H.}~\bibnamefont {Xiang}},\ }\href
  {https://doi.org/10.1103/physrevlett.120.197602} {\bibfield  {journal}
  {\bibinfo  {journal} {Phys. Rev. Lett.}\ }\textbf {\bibinfo {volume} {120}},\
  \bibinfo {pages} {97602} (\bibinfo {year} {2018})}\BibitemShut {NoStop}%
\bibitem [{\citenamefont {Aschauer}\ and\ \citenamefont
  {Spaldin}(2014)}]{Aschauer2014}%
  \BibitemOpen
  \bibfield  {author} {\bibinfo {author} {\bibfnamefont {U.}~\bibnamefont
  {Aschauer}}\ and\ \bibinfo {author} {\bibfnamefont {N.~A.}\ \bibnamefont
  {Spaldin}},\ }\href {https://doi.org/10.1088/0953-8984/26/12/122203}
  {\bibfield  {journal} {\bibinfo  {journal} {J. Phys. Condens. Matter}\
  }\textbf {\bibinfo {volume} {26}},\ \bibinfo {pages} {122203} (\bibinfo
  {year} {2014})}\BibitemShut {NoStop}%
\bibitem [{\citenamefont {Glazer}(1972)}]{Glazer1972}%
  \BibitemOpen
  \bibfield  {author} {\bibinfo {author} {\bibfnamefont {A.~M.}\ \bibnamefont
  {Glazer}},\ }\href {\doibase 10.1107/s0567740872007976} {\bibfield  {journal}
  {\bibinfo  {journal} {Acta Crystallogr. B}\ }\textbf {\bibinfo {volume}
  {28}},\ \bibinfo {pages} {3384} (\bibinfo {year} {1972})}\BibitemShut
  {NoStop}%
\bibitem [{\citenamefont {Glazer}(1975)}]{Glazer1975}%
  \BibitemOpen
  \bibfield  {author} {\bibinfo {author} {\bibfnamefont {A.~M.}\ \bibnamefont
  {Glazer}},\ }\href {\doibase 10.1107/s0567739475001635} {\bibfield  {journal}
  {\bibinfo  {journal} {Acta Crystallogr. A}\ }\textbf {\bibinfo {volume}
  {31}},\ \bibinfo {pages} {756} (\bibinfo {year} {1975})}\BibitemShut
  {NoStop}%
\bibitem [{\citenamefont {Munkholm}\ \emph {et~al.}(2001)\citenamefont
  {Munkholm}, \citenamefont {Streiffer}, \citenamefont {Murty}, \citenamefont
  {Eastman}, \citenamefont {Thompson}, \citenamefont {Auciello}, \citenamefont
  {Thompson}, \citenamefont {Moore},\ and\ \citenamefont
  {Stephenson}}]{Munkholm2001}%
  \BibitemOpen
  \bibfield  {author} {\bibinfo {author} {\bibfnamefont {A.}~\bibnamefont
  {Munkholm}}, \bibinfo {author} {\bibfnamefont {S.~K.}\ \bibnamefont
  {Streiffer}}, \bibinfo {author} {\bibfnamefont {M.~V.~R.}\ \bibnamefont
  {Murty}}, \bibinfo {author} {\bibfnamefont {J.~A.}\ \bibnamefont {Eastman}},
  \bibinfo {author} {\bibfnamefont {C.}~\bibnamefont {Thompson}}, \bibinfo
  {author} {\bibfnamefont {O.}~\bibnamefont {Auciello}}, \bibinfo {author}
  {\bibfnamefont {L.}~\bibnamefont {Thompson}}, \bibinfo {author}
  {\bibfnamefont {J.~F.}\ \bibnamefont {Moore}}, \ and\ \bibinfo {author}
  {\bibfnamefont {G.~B.}\ \bibnamefont {Stephenson}},\ }\href
  {https://doi.org/10.1103/physrevlett.88.016101} {\bibfield  {journal}
  {\bibinfo  {journal} {Phys. Rev. Lett.}\ }\textbf {\bibinfo {volume} {88}},\
  \bibinfo {pages} {016101} (\bibinfo {year} {2001})}\BibitemShut {NoStop}%
\bibitem [{\citenamefont {Bungaro}\ and\ \citenamefont
  {Rabe}(2005)}]{Bungaro2005}%
  \BibitemOpen
  \bibfield  {author} {\bibinfo {author} {\bibfnamefont {C.}~\bibnamefont
  {Bungaro}}\ and\ \bibinfo {author} {\bibfnamefont {K.~M.}\ \bibnamefont
  {Rabe}},\ }\href {https://doi.org/10.1103/physrevb.71.035420} {\bibfield
  {journal} {\bibinfo  {journal} {Phys. Rev. B}\ }\textbf {\bibinfo {volume}
  {71}},\ \bibinfo {pages} {035420} (\bibinfo {year} {2005})}\BibitemShut
  {NoStop}%
\bibitem [{\citenamefont {Sepliarsky}\ \emph {et~al.}(2005)\citenamefont
  {Sepliarsky}, \citenamefont {Stachiotti},\ and\ \citenamefont
  {Migoni}}]{Sepliarsky2005}%
  \BibitemOpen
  \bibfield  {author} {\bibinfo {author} {\bibfnamefont {M.}~\bibnamefont
  {Sepliarsky}}, \bibinfo {author} {\bibfnamefont {M.~G.}\ \bibnamefont
  {Stachiotti}}, \ and\ \bibinfo {author} {\bibfnamefont {R.~L.}\ \bibnamefont
  {Migoni}},\ }\href {\doibase 10.1103/physrevb.72.014110} {\bibfield
  {journal} {\bibinfo  {journal} {Phys. Rev. B}\ }\textbf {\bibinfo {volume}
  {72}},\ \bibinfo {pages} {014110} (\bibinfo {year} {2005})}\BibitemShut
  {NoStop}%
\bibitem [{\citenamefont {Bousquet}\ \emph {et~al.}(2008)\citenamefont
  {Bousquet}, \citenamefont {Dawber}, \citenamefont {Stucki}, \citenamefont
  {Lichtensteiger}, \citenamefont {Hermet}, \citenamefont {Gariglio},
  \citenamefont {Triscone},\ and\ \citenamefont {Ghosez}}]{Bousquet2008}%
  \BibitemOpen
  \bibfield  {author} {\bibinfo {author} {\bibfnamefont {E.}~\bibnamefont
  {Bousquet}}, \bibinfo {author} {\bibfnamefont {M.}~\bibnamefont {Dawber}},
  \bibinfo {author} {\bibfnamefont {N.}~\bibnamefont {Stucki}}, \bibinfo
  {author} {\bibfnamefont {C.}~\bibnamefont {Lichtensteiger}}, \bibinfo
  {author} {\bibfnamefont {P.}~\bibnamefont {Hermet}}, \bibinfo {author}
  {\bibfnamefont {S.}~\bibnamefont {Gariglio}}, \bibinfo {author}
  {\bibfnamefont {J.-M.}\ \bibnamefont {Triscone}}, \ and\ \bibinfo {author}
  {\bibfnamefont {P.}~\bibnamefont {Ghosez}},\ }\href
  {https://doi.org/10.1038/nature06817} {\bibfield  {journal} {\bibinfo
  {journal} {Nature}\ }\textbf {\bibinfo {volume} {452}},\ \bibinfo {pages}
  {732} (\bibinfo {year} {2008})}\BibitemShut {NoStop}%
\bibitem [{\citenamefont {Aguado-Puente}\ \emph {et~al.}(2011)\citenamefont
  {Aguado-Puente}, \citenamefont {Garc{\'{\i}}a-Fern{\'{a}}ndez},\ and\
  \citenamefont {Junquera}}]{AguadoPuente2011}%
  \BibitemOpen
  \bibfield  {author} {\bibinfo {author} {\bibfnamefont {P.}~\bibnamefont
  {Aguado-Puente}}, \bibinfo {author} {\bibfnamefont {P.}~\bibnamefont
  {Garc{\'{\i}}a-Fern{\'{a}}ndez}}, \ and\ \bibinfo {author} {\bibfnamefont
  {J.}~\bibnamefont {Junquera}},\ }\href
  {https://doi.org/10.1103/physrevlett.107.217601} {\bibfield  {journal}
  {\bibinfo  {journal} {Phys. Rev. Lett.}\ }\textbf {\bibinfo {volume} {107}},\
  \bibinfo {pages} {217601} (\bibinfo {year} {2011})}\BibitemShut {NoStop}%
\bibitem [{\citenamefont {Aguado-Puente}\ and\ \citenamefont
  {Junquera}(2012)}]{AguadoPuente2012}%
  \BibitemOpen
  \bibfield  {author} {\bibinfo {author} {\bibfnamefont {P.}~\bibnamefont
  {Aguado-Puente}}\ and\ \bibinfo {author} {\bibfnamefont {J.}~\bibnamefont
  {Junquera}},\ }\href {https://doi.org/10.1103/physrevb.85.184105} {\bibfield
  {journal} {\bibinfo  {journal} {Phys. Rev. B}\ }\textbf {\bibinfo {volume}
  {85}},\ \bibinfo {pages} {184105} (\bibinfo {year} {2012})}\BibitemShut
  {NoStop}%
\bibitem [{\citenamefont {Shimada}\ \emph {et~al.}(2010)\citenamefont
  {Shimada}, \citenamefont {Tomoda},\ and\ \citenamefont
  {Kitamura}}]{Shimada2010}%
  \BibitemOpen
  \bibfield  {author} {\bibinfo {author} {\bibfnamefont {T.}~\bibnamefont
  {Shimada}}, \bibinfo {author} {\bibfnamefont {S.}~\bibnamefont {Tomoda}}, \
  and\ \bibinfo {author} {\bibfnamefont {T.}~\bibnamefont {Kitamura}},\ }\href
  {https://doi.org/10.1103/physrevb.81.144116} {\bibfield  {journal} {\bibinfo
  {journal} {Phys. Rev. B}\ }\textbf {\bibinfo {volume} {81}},\ \bibinfo
  {pages} {144116} (\bibinfo {year} {2010})}\BibitemShut {NoStop}%
\bibitem [{\citenamefont {Junquera}\ and\ \citenamefont
  {Ghosez}(2003)}]{Junquera2003}%
  \BibitemOpen
  \bibfield  {author} {\bibinfo {author} {\bibfnamefont {J.}~\bibnamefont
  {Junquera}}\ and\ \bibinfo {author} {\bibfnamefont {P.}~\bibnamefont
  {Ghosez}},\ }\href {https://doi.org/10.1038/nature01501} {\bibfield
  {journal} {\bibinfo  {journal} {Nature}\ }\textbf {\bibinfo {volume} {422}},\
  \bibinfo {pages} {506} (\bibinfo {year} {2003})}\BibitemShut {NoStop}%
\bibitem [{\citenamefont {Aguado-Puente}\ \emph {et~al.}(2015)\citenamefont
  {Aguado-Puente}, \citenamefont {Bristowe}, \citenamefont {Yin}, \citenamefont
  {Shirasawa}, \citenamefont {Ghosez}, \citenamefont {Littlewood},\ and\
  \citenamefont {Artacho}}]{aguado2015model}%
  \BibitemOpen
  \bibfield  {author} {\bibinfo {author} {\bibfnamefont {P.}~\bibnamefont
  {Aguado-Puente}}, \bibinfo {author} {\bibfnamefont {N.~C.}\ \bibnamefont
  {Bristowe}}, \bibinfo {author} {\bibfnamefont {B.}~\bibnamefont {Yin}},
  \bibinfo {author} {\bibfnamefont {R.}~\bibnamefont {Shirasawa}}, \bibinfo
  {author} {\bibfnamefont {P.}~\bibnamefont {Ghosez}}, \bibinfo {author}
  {\bibfnamefont {P.~B.}\ \bibnamefont {Littlewood}}, \ and\ \bibinfo {author}
  {\bibfnamefont {E.}~\bibnamefont {Artacho}},\ }\href
  {https://doi.org/10.1103/physrevb.92.035438} {\bibfield  {journal} {\bibinfo
  {journal} {Phy. Rev. B}\ }\textbf {\bibinfo {volume} {92}},\ \bibinfo {pages}
  {035438} (\bibinfo {year} {2015})}\BibitemShut {NoStop}%
\bibitem [{\citenamefont {Streiffer}\ \emph {et~al.}(2002)\citenamefont
  {Streiffer}, \citenamefont {Eastman}, \citenamefont {Fong}, \citenamefont
  {Thompson}, \citenamefont {Munkholm}, \citenamefont {Murty}, \citenamefont
  {Auciello}, \citenamefont {Bai},\ and\ \citenamefont
  {Stephenson}}]{Streiffer2002}%
  \BibitemOpen
  \bibfield  {author} {\bibinfo {author} {\bibfnamefont {S.~K.}\ \bibnamefont
  {Streiffer}}, \bibinfo {author} {\bibfnamefont {J.~A.}\ \bibnamefont
  {Eastman}}, \bibinfo {author} {\bibfnamefont {D.~D.}\ \bibnamefont {Fong}},
  \bibinfo {author} {\bibfnamefont {C.}~\bibnamefont {Thompson}}, \bibinfo
  {author} {\bibfnamefont {A.}~\bibnamefont {Munkholm}}, \bibinfo {author}
  {\bibfnamefont {M.~V.~R.}\ \bibnamefont {Murty}}, \bibinfo {author}
  {\bibfnamefont {O.}~\bibnamefont {Auciello}}, \bibinfo {author}
  {\bibfnamefont {G.~R.}\ \bibnamefont {Bai}}, \ and\ \bibinfo {author}
  {\bibfnamefont {G.~B.}\ \bibnamefont {Stephenson}},\ }\href
  {https://doi.org/10.1103/physrevlett.89.067601} {\bibfield  {journal}
  {\bibinfo  {journal} {Phys. Rev. Lett.}\ }\textbf {\bibinfo {volume} {89}},\
  \bibinfo {pages} {067601} (\bibinfo {year} {2002})}\BibitemShut {NoStop}%
\bibitem [{\citenamefont {Fong}(2004)}]{Fong2004}%
  \BibitemOpen
  \bibfield  {author} {\bibinfo {author} {\bibfnamefont {D.~D.}\ \bibnamefont
  {Fong}},\ }\href {https://doi.org/10.1126/science.1098252} {\bibfield
  {journal} {\bibinfo  {journal} {Science}\ }\textbf {\bibinfo {volume}
  {304}},\ \bibinfo {pages} {1650} (\bibinfo {year} {2004})}\BibitemShut
  {NoStop}%
\bibitem [{\citenamefont {Bowler}\ \emph {et~al.}(2006)\citenamefont {Bowler},
  \citenamefont {Choudhury}, \citenamefont {Gillan},\ and\ \citenamefont
  {Miyazaki}}]{bowler2006recent}%
  \BibitemOpen
  \bibfield  {author} {\bibinfo {author} {\bibfnamefont {D.~R.}\ \bibnamefont
  {Bowler}}, \bibinfo {author} {\bibfnamefont {R.}~\bibnamefont {Choudhury}},
  \bibinfo {author} {\bibfnamefont {M.~J.}\ \bibnamefont {Gillan}}, \ and\
  \bibinfo {author} {\bibfnamefont {T.}~\bibnamefont {Miyazaki}},\ }\href
  {https://doi.org/10.1002/pssb.200541386} {\bibfield  {journal} {\bibinfo
  {journal} {Phys. Status Solidi. B}\ }\textbf {\bibinfo {volume} {243}},\
  \bibinfo {pages} {989} (\bibinfo {year} {2006})}\BibitemShut {NoStop}%
\bibitem [{\citenamefont {Bowler}\ and\ \citenamefont
  {Miyazaki}(2012)}]{bowler2012methods}%
  \BibitemOpen
  \bibfield  {author} {\bibinfo {author} {\bibfnamefont {D.}~\bibnamefont
  {Bowler}}\ and\ \bibinfo {author} {\bibfnamefont {T.}~\bibnamefont
  {Miyazaki}},\ }\href
  {http://iopscience.iop.org/article/10.1088/0034-4885/75/3/036503/meta}
  {\bibfield  {journal} {\bibinfo  {journal} {Rep. Prog. Phys.}\ }\textbf
  {\bibinfo {volume} {75}},\ \bibinfo {pages} {036503} (\bibinfo {year}
  {2012})}\BibitemShut {NoStop}%
\bibitem [{\citenamefont {{Ghosez}}\ and\ \citenamefont
  {{Junquera}}(2006)}]{ghosez2006first}%
  \BibitemOpen
  \bibfield  {author} {\bibinfo {author} {\bibfnamefont {P.}~\bibnamefont
  {{Ghosez}}}\ and\ \bibinfo {author} {\bibfnamefont {J.}~\bibnamefont
  {{Junquera}}},\ }\href {http://adsabs.harvard.edu/abs/2006cond.mat..5299G}
  {\bibfield  {journal} {\bibinfo  {journal} {eprint arXiv:cond-mat/0605299}\ }
  (\bibinfo {year} {2006})},\ \Eprint {http://arxiv.org/abs/cond-mat/0605299}
  {cond-mat/0605299} \BibitemShut {NoStop}%
\bibitem [{\citenamefont {Garc{\'{\i}}a-Fern{\'{a}}ndez}\ \emph
  {et~al.}(2016)\citenamefont {Garc{\'{\i}}a-Fern{\'{a}}ndez}, \citenamefont
  {Wojde{\l}}, \citenamefont {{\'{I}}{\~{n}}iguez},\ and\ \citenamefont
  {Junquera}}]{garcia2016second}%
  \BibitemOpen
  \bibfield  {author} {\bibinfo {author} {\bibfnamefont {P.}~\bibnamefont
  {Garc{\'{\i}}a-Fern{\'{a}}ndez}}, \bibinfo {author} {\bibfnamefont {J.~C.}\
  \bibnamefont {Wojde{\l}}}, \bibinfo {author} {\bibfnamefont {J.}~\bibnamefont
  {{\'{I}}{\~{n}}iguez}}, \ and\ \bibinfo {author} {\bibfnamefont
  {J.}~\bibnamefont {Junquera}},\ }\href
  {https://doi.org/10.1103/physrevb.93.195137} {\bibfield  {journal} {\bibinfo
  {journal} {Phys. Rev. B}\ }\textbf {\bibinfo {volume} {93}},\ \bibinfo
  {pages} {195137} (\bibinfo {year} {2016})}\BibitemShut {NoStop}%
\bibitem [{\citenamefont {Chen}(2002)}]{chen2002phase}%
  \BibitemOpen
  \bibfield  {author} {\bibinfo {author} {\bibfnamefont {L.-Q.}\ \bibnamefont
  {Chen}},\ }\href {https://doi.org/10.1146/annurev.matsci.32.112001.132041}
  {\bibfield  {journal} {\bibinfo  {journal} {Annu. Rev. Mater. Res.}\ }\textbf
  {\bibinfo {volume} {32}},\ \bibinfo {pages} {113} (\bibinfo {year}
  {2002})}\BibitemShut {NoStop}%
\bibitem [{\citenamefont {Shafer}\ \emph {et~al.}(2018)\citenamefont {Shafer},
  \citenamefont {Garc{\'{\i}}a-Fern{\'{a}}ndez}, \citenamefont {Aguado-Puente},
  \citenamefont {Damodaran}, \citenamefont {Yadav}, \citenamefont {Nelson},
  \citenamefont {Hsu}, \citenamefont {Wojde{\l}}, \citenamefont
  {{\'{I}}{\~{n}}iguez}, \citenamefont {Martin}, \citenamefont {Arenholz},
  \citenamefont {Junquera},\ and\ \citenamefont {Ramesh}}]{Shafer2018}%
  \BibitemOpen
  \bibfield  {author} {\bibinfo {author} {\bibfnamefont {P.}~\bibnamefont
  {Shafer}}, \bibinfo {author} {\bibfnamefont {P.}~\bibnamefont
  {Garc{\'{\i}}a-Fern{\'{a}}ndez}}, \bibinfo {author} {\bibfnamefont
  {P.}~\bibnamefont {Aguado-Puente}}, \bibinfo {author} {\bibfnamefont {A.~R.}\
  \bibnamefont {Damodaran}}, \bibinfo {author} {\bibfnamefont {A.~K.}\
  \bibnamefont {Yadav}}, \bibinfo {author} {\bibfnamefont {C.~T.}\ \bibnamefont
  {Nelson}}, \bibinfo {author} {\bibfnamefont {S.-L.}\ \bibnamefont {Hsu}},
  \bibinfo {author} {\bibfnamefont {J.~C.}\ \bibnamefont {Wojde{\l}}}, \bibinfo
  {author} {\bibfnamefont {J.}~\bibnamefont {{\'{I}}{\~{n}}iguez}}, \bibinfo
  {author} {\bibfnamefont {L.~W.}\ \bibnamefont {Martin}}, \bibinfo {author}
  {\bibfnamefont {E.}~\bibnamefont {Arenholz}}, \bibinfo {author}
  {\bibfnamefont {J.}~\bibnamefont {Junquera}}, \ and\ \bibinfo {author}
  {\bibfnamefont {R.}~\bibnamefont {Ramesh}},\ }\href
  {https://doi.org/10.1073/pnas.1711652115} {\bibfield  {journal} {\bibinfo
  {journal} {Proc. Natl. Acad. Sci. U.S.A.}\ }\textbf {\bibinfo {volume}
  {115}},\ \bibinfo {pages} {915} (\bibinfo {year} {2018})}\BibitemShut
  {NoStop}%
\bibitem [{\citenamefont {Chapman}\ \emph {et~al.}(2017)\citenamefont
  {Chapman}, \citenamefont {Kimmel},\ and\ \citenamefont
  {Duffy}}]{Chapman2017}%
  \BibitemOpen
  \bibfield  {author} {\bibinfo {author} {\bibfnamefont {J.~B.~J.}\
  \bibnamefont {Chapman}}, \bibinfo {author} {\bibfnamefont {A.~V.}\
  \bibnamefont {Kimmel}}, \ and\ \bibinfo {author} {\bibfnamefont {D.~M.}\
  \bibnamefont {Duffy}},\ }\href {\doibase 10.1039/c6cp08157f} {\bibfield
  {journal} {\bibinfo  {journal} {Phys. Chem. Chem. Phys.}\ }\textbf {\bibinfo
  {volume} {19}},\ \bibinfo {pages} {4243} (\bibinfo {year}
  {2017})}\BibitemShut {NoStop}%
\bibitem [{\citenamefont {Sepliarsky}\ \emph {et~al.}(2006)\citenamefont
  {Sepliarsky}, \citenamefont {Stachiotti},\ and\ \citenamefont
  {Migoni}}]{Sepliarsky2006}%
  \BibitemOpen
  \bibfield  {author} {\bibinfo {author} {\bibfnamefont {M.}~\bibnamefont
  {Sepliarsky}}, \bibinfo {author} {\bibfnamefont {M.~G.}\ \bibnamefont
  {Stachiotti}}, \ and\ \bibinfo {author} {\bibfnamefont {R.~L.}\ \bibnamefont
  {Migoni}},\ }\href {\doibase 10.1103/physrevlett.96.137603} {\bibfield
  {journal} {\bibinfo  {journal} {Phys. Rev. Lett.}\ }\textbf {\bibinfo
  {volume} {96}},\ \bibinfo {pages} {137603} (\bibinfo {year}
  {2006})}\BibitemShut {NoStop}%
\bibitem [{\citenamefont {Kohn}\ and\ \citenamefont {Sham}(1965)}]{Kohn1965}%
  \BibitemOpen
  \bibfield  {author} {\bibinfo {author} {\bibfnamefont {W.}~\bibnamefont
  {Kohn}}\ and\ \bibinfo {author} {\bibfnamefont {L.~J.}\ \bibnamefont
  {Sham}},\ }\href {\doibase 10.1103/physrev.140.a1133} {\bibfield  {journal}
  {\bibinfo  {journal} {Phys. Rev.}\ }\textbf {\bibinfo {volume} {140}},\
  \bibinfo {pages} {A1133} (\bibinfo {year} {1965})}\BibitemShut {NoStop}%
\bibitem [{\citenamefont {Hohenberg}\ and\ \citenamefont
  {Kohn}(1964)}]{Hohenberg1964}%
  \BibitemOpen
  \bibfield  {author} {\bibinfo {author} {\bibfnamefont {P.}~\bibnamefont
  {Hohenberg}}\ and\ \bibinfo {author} {\bibfnamefont {W.}~\bibnamefont
  {Kohn}},\ }\href {\doibase 10.1103/physrev.136.b864} {\bibfield  {journal}
  {\bibinfo  {journal} {Phys. Rev.}\ }\textbf {\bibinfo {volume} {136}},\
  \bibinfo {pages} {B864} (\bibinfo {year} {1964})}\BibitemShut {NoStop}%
\bibitem [{\citenamefont {Bowler}\ \emph {et~al.}(2020)\citenamefont {Bowler},
  \citenamefont {Baker}, \citenamefont {Poulton}, \citenamefont {Mujahed},
  \citenamefont {Lin}, \citenamefont {Yadav}, \citenamefont {Raza},
  \citenamefont {Miyazaki}, \citenamefont {Gillan}, \citenamefont {Nakata},
  \citenamefont {Truflandier}, \citenamefont {Torralba}, \citenamefont
  {Brazdova}, \citenamefont {Tong}, \citenamefont {Arita}, \citenamefont
  {Sena}, \citenamefont {Terranova}, \citenamefont {Choudhury}, \citenamefont
  {Goringe}, \citenamefont {Hernandez},\ and\ \citenamefont
  {Bush}}]{CQRelease2020}%
  \BibitemOpen
  \bibfield  {author} {\bibinfo {author} {\bibfnamefont {D.~R.}\ \bibnamefont
  {Bowler}}, \bibinfo {author} {\bibfnamefont {J.~S.}\ \bibnamefont {Baker}},
  \bibinfo {author} {\bibfnamefont {J.~T.~L.}\ \bibnamefont {Poulton}},
  \bibinfo {author} {\bibfnamefont {S.~Y.}\ \bibnamefont {Mujahed}}, \bibinfo
  {author} {\bibfnamefont {J.}~\bibnamefont {Lin}}, \bibinfo {author}
  {\bibfnamefont {S.}~\bibnamefont {Yadav}}, \bibinfo {author} {\bibfnamefont
  {Z.}~\bibnamefont {Raza}}, \bibinfo {author} {\bibfnamefont {T.}~\bibnamefont
  {Miyazaki}}, \bibinfo {author} {\bibfnamefont {M.}~\bibnamefont {Gillan}},
  \bibinfo {author} {\bibfnamefont {A.}~\bibnamefont {Nakata}}, \bibinfo
  {author} {\bibfnamefont {L.}~\bibnamefont {Truflandier}}, \bibinfo {author}
  {\bibfnamefont {A.}~\bibnamefont {Torralba}}, \bibinfo {author}
  {\bibfnamefont {V.}~\bibnamefont {Brazdova}}, \bibinfo {author}
  {\bibfnamefont {L.}~\bibnamefont {Tong}}, \bibinfo {author} {\bibfnamefont
  {M.}~\bibnamefont {Arita}}, \bibinfo {author} {\bibfnamefont
  {A.}~\bibnamefont {Sena}}, \bibinfo {author} {\bibfnamefont {U.}~\bibnamefont
  {Terranova}}, \bibinfo {author} {\bibfnamefont {R.}~\bibnamefont
  {Choudhury}}, \bibinfo {author} {\bibfnamefont {C.}~\bibnamefont {Goringe}},
  \bibinfo {author} {\bibfnamefont {E.}~\bibnamefont {Hernandez}}, \ and\
  \bibinfo {author} {\bibfnamefont {I.}~\bibnamefont {Bush}},\ }\href
  {https://github.com/OrderN/CONQUEST-release} {\enquote {\bibinfo {title} {The
  \textsc{CONQUEST} code: public release at
  \href{https://github.com/OrderN/CONQUEST-release}{https://github.com/OrderN/CONQUEST-release}},}\
  } (\bibinfo {year} {2020})\BibitemShut {NoStop}%
\bibitem [{\citenamefont {Bowler}\ and\ \citenamefont
  {Miyazaki}(2010)}]{Bowler2010}%
  \BibitemOpen
  \bibfield  {author} {\bibinfo {author} {\bibfnamefont {D.~R.}\ \bibnamefont
  {Bowler}}\ and\ \bibinfo {author} {\bibfnamefont {T.}~\bibnamefont
  {Miyazaki}},\ }\href {\doibase 10.1088/0953-8984/22/7/074207} {\bibfield
  {journal} {\bibinfo  {journal} {J. Phys. Condens. Matter}\ }\textbf {\bibinfo
  {volume} {22}},\ \bibinfo {pages} {074207} (\bibinfo {year}
  {2010})}\BibitemShut {NoStop}%
\bibitem [{\citenamefont {Rayson}\ and\ \citenamefont
  {Briddon}(2009)}]{rayson2009highly}%
  \BibitemOpen
  \bibfield  {author} {\bibinfo {author} {\bibfnamefont {M.~J.}\ \bibnamefont
  {Rayson}}\ and\ \bibinfo {author} {\bibfnamefont {P.~R.}\ \bibnamefont
  {Briddon}},\ }\href {https://doi.org/10.1103/physrevb.80.205104} {\bibfield
  {journal} {\bibinfo  {journal} {Phys. Rev. B}\ }\textbf {\bibinfo {volume}
  {80}},\ \bibinfo {pages} {05104} (\bibinfo {year} {2009})}\BibitemShut
  {NoStop}%
\bibitem [{\citenamefont {Bowler}\ \emph {et~al.}(2019)\citenamefont {Bowler},
  \citenamefont {Baker}, \citenamefont {Poulton}, \citenamefont {Mujahed},
  \citenamefont {Lin}, \citenamefont {Yadav}, \citenamefont {Raza},\ and\
  \citenamefont {Miyazaki}}]{Bowler2019}%
  \BibitemOpen
  \bibfield  {author} {\bibinfo {author} {\bibfnamefont {D.~R.}\ \bibnamefont
  {Bowler}}, \bibinfo {author} {\bibfnamefont {J.~S.}\ \bibnamefont {Baker}},
  \bibinfo {author} {\bibfnamefont {J.~T.~L.}\ \bibnamefont {Poulton}},
  \bibinfo {author} {\bibfnamefont {S.~Y.}\ \bibnamefont {Mujahed}}, \bibinfo
  {author} {\bibfnamefont {J.}~\bibnamefont {Lin}}, \bibinfo {author}
  {\bibfnamefont {S.}~\bibnamefont {Yadav}}, \bibinfo {author} {\bibfnamefont
  {Z.}~\bibnamefont {Raza}}, \ and\ \bibinfo {author} {\bibfnamefont
  {T.}~\bibnamefont {Miyazaki}},\ }\href {\doibase 10.7567/1347-4065/ab45af}
  {\bibfield  {journal} {\bibinfo  {journal} {Jpn. J. Appl. Phys.}\ }\textbf
  {\bibinfo {volume} {58}},\ \bibinfo {pages} {100503} (\bibinfo {year}
  {2019})}\BibitemShut {NoStop}%
\bibitem [{\citenamefont {Baker}\ \emph {et~al.}(2020)\citenamefont {Baker},
  \citenamefont {Miyazaki},\ and\ \citenamefont
  {Bowler}}]{Baker2020pseudoatomic}%
  \BibitemOpen
  \bibfield  {author} {\bibinfo {author} {\bibfnamefont {J.~S.}\ \bibnamefont
  {Baker}}, \bibinfo {author} {\bibfnamefont {T.}~\bibnamefont {Miyazaki}}, \
  and\ \bibinfo {author} {\bibfnamefont {D.~R.}\ \bibnamefont {Bowler}},\
  }\href {\doibase 10.1088/2516-1075/ab950e} {\bibfield  {journal} {\bibinfo
  {journal} {Electron. Struct.}\ } (\bibinfo {year} {2020}),\
  10.1088/2516-1075/ab950e}\BibitemShut {NoStop}%
\bibitem [{\citenamefont {Soler}\ \emph {et~al.}(2002)\citenamefont {Soler},
  \citenamefont {Artacho}, \citenamefont {Gale}, \citenamefont {Garc{\'{\i}}a},
  \citenamefont {Junquera}, \citenamefont {Ordej{\'{o}}n},\ and\ \citenamefont
  {S{\'{a}}nchez-Portal}}]{soler2002siesta}%
  \BibitemOpen
  \bibfield  {author} {\bibinfo {author} {\bibfnamefont {J.~M.}\ \bibnamefont
  {Soler}}, \bibinfo {author} {\bibfnamefont {E.}~\bibnamefont {Artacho}},
  \bibinfo {author} {\bibfnamefont {J.~D.}\ \bibnamefont {Gale}}, \bibinfo
  {author} {\bibfnamefont {A.}~\bibnamefont {Garc{\'{\i}}a}}, \bibinfo {author}
  {\bibfnamefont {J.}~\bibnamefont {Junquera}}, \bibinfo {author}
  {\bibfnamefont {P.}~\bibnamefont {Ordej{\'{o}}n}}, \ and\ \bibinfo {author}
  {\bibfnamefont {D.}~\bibnamefont {S{\'{a}}nchez-Portal}},\ }\href {\doibase
  10.1088/0953-8984/14/11/302} {\bibfield  {journal} {\bibinfo  {journal} {J.
  Phys. Condens. Matter}\ }\textbf {\bibinfo {volume} {14}},\ \bibinfo {pages}
  {2745} (\bibinfo {year} {2002})}\BibitemShut {NoStop}%
\bibitem [{\citenamefont {Torralba}\ \emph {et~al.}(2008)\citenamefont
  {Torralba}, \citenamefont {Todorovi{\'c}}, \citenamefont {Br{\'a}zdov{\'a}},
  \citenamefont {Choudhury}, \citenamefont {Miyazaki}, \citenamefont {Gillan},\
  and\ \citenamefont {Bowler}}]{torralba2008pseudo}%
  \BibitemOpen
  \bibfield  {author} {\bibinfo {author} {\bibfnamefont {A.}~\bibnamefont
  {Torralba}}, \bibinfo {author} {\bibfnamefont {M.}~\bibnamefont
  {Todorovi{\'c}}}, \bibinfo {author} {\bibfnamefont {V.}~\bibnamefont
  {Br{\'a}zdov{\'a}}}, \bibinfo {author} {\bibfnamefont {R.}~\bibnamefont
  {Choudhury}}, \bibinfo {author} {\bibfnamefont {T.}~\bibnamefont {Miyazaki}},
  \bibinfo {author} {\bibfnamefont {M.}~\bibnamefont {Gillan}}, \ and\ \bibinfo
  {author} {\bibfnamefont {D.}~\bibnamefont {Bowler}},\ }\href
  {http://stacks.iop.org/0953-8984/20/i=29/a=294206} {\bibfield  {journal}
  {\bibinfo  {journal} {J. Phys. Condens. Matter}\ }\textbf {\bibinfo {volume}
  {20}},\ \bibinfo {pages} {294206} (\bibinfo {year} {2008})}\BibitemShut
  {NoStop}%
\bibitem [{\citenamefont {Rayson}(2010)}]{rayson2010rapid}%
  \BibitemOpen
  \bibfield  {author} {\bibinfo {author} {\bibfnamefont {M.}~\bibnamefont
  {Rayson}},\ }\href {https://doi.org/10.1016/j.cpc.2010.02.012} {\bibfield
  {journal} {\bibinfo  {journal} {Comput. Phys. Commun.}\ }\textbf {\bibinfo
  {volume} {181}},\ \bibinfo {pages} {1051} (\bibinfo {year}
  {2010})}\BibitemShut {NoStop}%
\bibitem [{\citenamefont {Perdew}\ and\ \citenamefont
  {Zunger}(1981)}]{perdew1981self}%
  \BibitemOpen
  \bibfield  {author} {\bibinfo {author} {\bibfnamefont {J.~P.}\ \bibnamefont
  {Perdew}}\ and\ \bibinfo {author} {\bibfnamefont {A.}~\bibnamefont
  {Zunger}},\ }\href {https://doi.org/10.1103/physrevb.23.5048} {\bibfield
  {journal} {\bibinfo  {journal} {Phys. Rev. B}\ }\textbf {\bibinfo {volume}
  {23}},\ \bibinfo {pages} {5048} (\bibinfo {year} {1981})}\BibitemShut
  {NoStop}%
\bibitem [{\citenamefont {Ceperley}\ and\ \citenamefont
  {Alder}(1980)}]{ceperley1980ground}%
  \BibitemOpen
  \bibfield  {author} {\bibinfo {author} {\bibfnamefont {D.~M.}\ \bibnamefont
  {Ceperley}}\ and\ \bibinfo {author} {\bibfnamefont {B.~J.}\ \bibnamefont
  {Alder}},\ }\href {https://doi.org/10.1103/physrevlett.45.566} {\bibfield
  {journal} {\bibinfo  {journal} {Phys. Rev. Lett.}\ }\textbf {\bibinfo
  {volume} {45}},\ \bibinfo {pages} {566} (\bibinfo {year} {1980})}\BibitemShut
  {NoStop}%
\bibitem [{\citenamefont {Zhang}\ \emph {et~al.}(2017)\citenamefont {Zhang},
  \citenamefont {Sun}, \citenamefont {Perdew},\ and\ \citenamefont
  {Wu}}]{Zhang2017}%
  \BibitemOpen
  \bibfield  {author} {\bibinfo {author} {\bibfnamefont {Y.}~\bibnamefont
  {Zhang}}, \bibinfo {author} {\bibfnamefont {J.}~\bibnamefont {Sun}}, \bibinfo
  {author} {\bibfnamefont {J.~P.}\ \bibnamefont {Perdew}}, \ and\ \bibinfo
  {author} {\bibfnamefont {X.}~\bibnamefont {Wu}},\ }\href
  {https://doi.org/10.1103/physrevb.96.035143} {\bibfield  {journal} {\bibinfo
  {journal} {Phys. Rev. B}\ }\textbf {\bibinfo {volume} {96}},\ \bibinfo
  {pages} {035143} (\bibinfo {year} {2017})}\BibitemShut {NoStop}%
\bibitem [{\citenamefont {Hamann}(2013)}]{hamann2013optimized}%
  \BibitemOpen
  \bibfield  {author} {\bibinfo {author} {\bibfnamefont {D.~R.}\ \bibnamefont
  {Hamann}},\ }\href {https://doi.org/10.1103/physrevb.88.085117} {\bibfield
  {journal} {\bibinfo  {journal} {Phys. Rev. B}\ }\textbf {\bibinfo {volume}
  {88}},\ \bibinfo {pages} {085117} (\bibinfo {year} {2013})}\BibitemShut
  {NoStop}%
\bibitem [{\citenamefont {Vanderbilt}(1990)}]{Vanderbilt1990}%
  \BibitemOpen
  \bibfield  {author} {\bibinfo {author} {\bibfnamefont {D.}~\bibnamefont
  {Vanderbilt}},\ }\href {https://doi.org/10.1103/physrevb.41.7892} {\bibfield
  {journal} {\bibinfo  {journal} {Phys. Rev. B}\ }\textbf {\bibinfo {volume}
  {41}},\ \bibinfo {pages} {7892} (\bibinfo {year} {1990})}\BibitemShut
  {NoStop}%
\bibitem [{\citenamefont {van Setten}\ \emph {et~al.}(2018)\citenamefont {van
  Setten}, \citenamefont {Giantomassi}, \citenamefont {Bousquet}, \citenamefont
  {Verstraete}, \citenamefont {Hamann}, \citenamefont {Gonze},\ and\
  \citenamefont {Rignanese}}]{van2018pseudodojo}%
  \BibitemOpen
  \bibfield  {author} {\bibinfo {author} {\bibfnamefont {M.}~\bibnamefont {van
  Setten}}, \bibinfo {author} {\bibfnamefont {M.}~\bibnamefont {Giantomassi}},
  \bibinfo {author} {\bibfnamefont {E.}~\bibnamefont {Bousquet}}, \bibinfo
  {author} {\bibfnamefont {M.}~\bibnamefont {Verstraete}}, \bibinfo {author}
  {\bibfnamefont {D.}~\bibnamefont {Hamann}}, \bibinfo {author} {\bibfnamefont
  {X.}~\bibnamefont {Gonze}}, \ and\ \bibinfo {author} {\bibfnamefont {G.-M.}\
  \bibnamefont {Rignanese}},\ }\href
  {https://doi.org/10.1016/j.cpc.2018.01.012} {\bibfield  {journal} {\bibinfo
  {journal} {Comput. Phys. Commun.}\ }\textbf {\bibinfo {volume} {226}},\
  \bibinfo {pages} {39} (\bibinfo {year} {2018})}\BibitemShut {NoStop}%
\bibitem [{\citenamefont {Monkhorst}\ and\ \citenamefont
  {Pack}(1976)}]{monkhorst1976special}%
  \BibitemOpen
  \bibfield  {author} {\bibinfo {author} {\bibfnamefont {H.~J.}\ \bibnamefont
  {Monkhorst}}\ and\ \bibinfo {author} {\bibfnamefont {J.~D.}\ \bibnamefont
  {Pack}},\ }\href {https://doi.org/10.1103/physrevb.13.5188} {\bibfield
  {journal} {\bibinfo  {journal} {Phys. Rev. B}\ }\textbf {\bibinfo {volume}
  {13}},\ \bibinfo {pages} {5188} (\bibinfo {year} {1976})}\BibitemShut
  {NoStop}%
\bibitem [{\citenamefont {Resta}(1999)}]{Resta1999}%
  \BibitemOpen
  \bibfield  {author} {\bibinfo {author} {\bibfnamefont {R.}~\bibnamefont
  {Resta}},\ }\href {\doibase
  10.1002/(sici)1097-461x(1999)75:4/5<599::aid-qua25>3.0.co;2-8} {\bibfield
  {journal} {\bibinfo  {journal} {Int. J. Quantum Chem.}\ }\textbf {\bibinfo
  {volume} {75}},\ \bibinfo {pages} {599} (\bibinfo {year} {1999})}\BibitemShut
  {NoStop}%
\bibitem [{\citenamefont {Resta}\ \emph {et~al.}(1993)\citenamefont {Resta},
  \citenamefont {Posternak},\ and\ \citenamefont {Baldereschi}}]{Resta1993}%
  \BibitemOpen
  \bibfield  {author} {\bibinfo {author} {\bibfnamefont {R.}~\bibnamefont
  {Resta}}, \bibinfo {author} {\bibfnamefont {M.}~\bibnamefont {Posternak}}, \
  and\ \bibinfo {author} {\bibfnamefont {A.}~\bibnamefont {Baldereschi}},\
  }\href {\doibase 10.1103/physrevlett.70.1010} {\bibfield  {journal} {\bibinfo
   {journal} {Phys. Rev. Lett}\ }\textbf {\bibinfo {volume} {70}},\ \bibinfo
  {pages} {1010} (\bibinfo {year} {1993})}\BibitemShut {NoStop}%
\bibitem [{\citenamefont {Berry}(1984)}]{Berry1984}%
  \BibitemOpen
  \bibfield  {author} {\bibinfo {author} {\bibfnamefont {M.~V.}\ \bibnamefont
  {Berry}},\ }\href {\doibase 10.1098/rspa.1984.0023} {\bibfield  {journal}
  {\bibinfo  {journal} {Proc. R. Soc. A}\ }\textbf {\bibinfo {volume} {392}},\
  \bibinfo {pages} {45} (\bibinfo {year} {1984})}\BibitemShut {NoStop}%
\bibitem [{\citenamefont {King-Smith}\ and\ \citenamefont
  {Vanderbilt}(1993)}]{KingSmith1993}%
  \BibitemOpen
  \bibfield  {author} {\bibinfo {author} {\bibfnamefont {R.~D.}\ \bibnamefont
  {King-Smith}}\ and\ \bibinfo {author} {\bibfnamefont {D.}~\bibnamefont
  {Vanderbilt}},\ }\href {\doibase 10.1103/physrevb.47.1651} {\bibfield
  {journal} {\bibinfo  {journal} {Phys. Rev. B.}\ }\textbf {\bibinfo {volume}
  {47}},\ \bibinfo {pages} {1651} (\bibinfo {year} {1993})}\BibitemShut
  {NoStop}%
\bibitem [{\citenamefont {Bengtsson}(1999)}]{bengtsson1999dipole}%
  \BibitemOpen
  \bibfield  {author} {\bibinfo {author} {\bibfnamefont {L.}~\bibnamefont
  {Bengtsson}},\ }\href {https://doi.org/10.1103/physrevb.59.12301} {\bibfield
  {journal} {\bibinfo  {journal} {Phys. Rev. B}\ }\textbf {\bibinfo {volume}
  {59}},\ \bibinfo {pages} {12301} (\bibinfo {year} {1999})}\BibitemShut
  {NoStop}%
\bibitem [{\citenamefont {Lebedev}(2009)}]{lebedev2009ab}%
  \BibitemOpen
  \bibfield  {author} {\bibinfo {author} {\bibfnamefont {A.~I.}\ \bibnamefont
  {Lebedev}},\ }\href {https://doi.org/10.1134/s1063783409020279} {\bibfield
  {journal} {\bibinfo  {journal} {Phys. Solid State}\ }\textbf {\bibinfo
  {volume} {51}},\ \bibinfo {pages} {362} (\bibinfo {year} {2009})}\BibitemShut
  {NoStop}%
\bibitem [{\citenamefont {Lytle}(1964)}]{Lytle1964}%
  \BibitemOpen
  \bibfield  {author} {\bibinfo {author} {\bibfnamefont {F.~W.}\ \bibnamefont
  {Lytle}},\ }\href {https://doi.org/10.1063/1.1702820} {\bibfield  {journal}
  {\bibinfo  {journal} {J. Appl. Phys.}\ }\textbf {\bibinfo {volume} {35}},\
  \bibinfo {pages} {2212} (\bibinfo {year} {1964})}\BibitemShut {NoStop}%
\bibitem [{\citenamefont {Heifets}\ \emph {et~al.}(2006)\citenamefont
  {Heifets}, \citenamefont {Kotomin},\ and\ \citenamefont
  {Trepakov}}]{heifets2006calculations}%
  \BibitemOpen
  \bibfield  {author} {\bibinfo {author} {\bibfnamefont {E.}~\bibnamefont
  {Heifets}}, \bibinfo {author} {\bibfnamefont {E.}~\bibnamefont {Kotomin}}, \
  and\ \bibinfo {author} {\bibfnamefont {V.}~\bibnamefont {Trepakov}},\ }\href
  {http://stacks.iop.org/0953-8984/18/i=20/a=009} {\bibfield  {journal}
  {\bibinfo  {journal} {J. Phys. Condens. Matter}\ }\textbf {\bibinfo {volume}
  {18}},\ \bibinfo {pages} {4845} (\bibinfo {year} {2006})}\BibitemShut
  {NoStop}%
\bibitem [{\citenamefont {Umeno}\ \emph {et~al.}(2006)\citenamefont {Umeno},
  \citenamefont {Shimada}, \citenamefont {Kitamura},\ and\ \citenamefont
  {Els\"{a}sser}}]{Umeno2006}%
  \BibitemOpen
  \bibfield  {author} {\bibinfo {author} {\bibfnamefont {Y.}~\bibnamefont
  {Umeno}}, \bibinfo {author} {\bibfnamefont {T.}~\bibnamefont {Shimada}},
  \bibinfo {author} {\bibfnamefont {T.}~\bibnamefont {Kitamura}}, \ and\
  \bibinfo {author} {\bibfnamefont {C.}~\bibnamefont {Els\"{a}sser}},\ }\href
  {\doibase 10.1103/physrevb.74.174111} {\bibfield  {journal} {\bibinfo
  {journal} {Physical Review B}\ }\textbf {\bibinfo {volume} {74}},\ \bibinfo
  {pages} {174111} (\bibinfo {year} {2006})}\BibitemShut {NoStop}%
\bibitem [{\citenamefont {Meyer}\ and\ \citenamefont
  {Vanderbilt}(2002)}]{Meyer2002}%
  \BibitemOpen
  \bibfield  {author} {\bibinfo {author} {\bibfnamefont {B.}~\bibnamefont
  {Meyer}}\ and\ \bibinfo {author} {\bibfnamefont {D.}~\bibnamefont
  {Vanderbilt}},\ }\href {\doibase 10.1103/physrevb.65.104111} {\bibfield
  {journal} {\bibinfo  {journal} {Phys. Rev. B}\ }\textbf {\bibinfo {volume}
  {65}},\ \bibinfo {pages} {104111} (\bibinfo {year} {2002})}\BibitemShut
  {NoStop}%
\bibitem [{\citenamefont {Glinchuk}\ \emph {et~al.}(2010)\citenamefont
  {Glinchuk}, \citenamefont {Morozovska},\ and\ \citenamefont
  {Eliseev}}]{Glinchuk2010}%
  \BibitemOpen
  \bibfield  {author} {\bibinfo {author} {\bibfnamefont {M.~D.}\ \bibnamefont
  {Glinchuk}}, \bibinfo {author} {\bibfnamefont {A.~N.}\ \bibnamefont
  {Morozovska}}, \ and\ \bibinfo {author} {\bibfnamefont {E.~A.}\ \bibnamefont
  {Eliseev}},\ }\href {\doibase 10.1080/00150193.2010.505796} {\bibfield
  {journal} {\bibinfo  {journal} {Ferroelectrics}\ }\textbf {\bibinfo {volume}
  {400}},\ \bibinfo {pages} {243} (\bibinfo {year} {2010})}\BibitemShut
  {NoStop}%
\bibitem [{\citenamefont {Sai}\ \emph {et~al.}(2000)\citenamefont {Sai},
  \citenamefont {Meyer},\ and\ \citenamefont {Vanderbilt}}]{Sai2000}%
  \BibitemOpen
  \bibfield  {author} {\bibinfo {author} {\bibfnamefont {N.}~\bibnamefont
  {Sai}}, \bibinfo {author} {\bibfnamefont {B.}~\bibnamefont {Meyer}}, \ and\
  \bibinfo {author} {\bibfnamefont {D.}~\bibnamefont {Vanderbilt}},\ }\href
  {\doibase 10.1103/physrevlett.84.5636} {\bibfield  {journal} {\bibinfo
  {journal} {Phys. Rev. Lett.}\ }\textbf {\bibinfo {volume} {84}},\ \bibinfo
  {pages} {5636} (\bibinfo {year} {2000})}\BibitemShut {NoStop}%
\bibitem [{\citenamefont {Warusawithana}\ \emph {et~al.}(2003)\citenamefont
  {Warusawithana}, \citenamefont {Colla}, \citenamefont {Eckstein},\ and\
  \citenamefont {Weissman}}]{Warusawithana2003}%
  \BibitemOpen
  \bibfield  {author} {\bibinfo {author} {\bibfnamefont {M.~P.}\ \bibnamefont
  {Warusawithana}}, \bibinfo {author} {\bibfnamefont {E.~V.}\ \bibnamefont
  {Colla}}, \bibinfo {author} {\bibfnamefont {J.~N.}\ \bibnamefont {Eckstein}},
  \ and\ \bibinfo {author} {\bibfnamefont {M.~B.}\ \bibnamefont {Weissman}},\
  }\href {\doibase 10.1103/physrevlett.90.036802} {\bibfield  {journal}
  {\bibinfo  {journal} {Phys. Rev. Lett.}\ }\textbf {\bibinfo {volume} {90}}
  (\bibinfo {year} {2003}),\ 10.1103/physrevlett.90.036802}\BibitemShut
  {NoStop}%
\bibitem [{\citenamefont {Lichtensteiger}\ \emph {et~al.}(2005)\citenamefont
  {Lichtensteiger}, \citenamefont {Triscone}, \citenamefont {Junquera},\ and\
  \citenamefont {Ghosez}}]{Lichtensteiger2005}%
  \BibitemOpen
  \bibfield  {author} {\bibinfo {author} {\bibfnamefont {C.}~\bibnamefont
  {Lichtensteiger}}, \bibinfo {author} {\bibfnamefont {J.-M.}\ \bibnamefont
  {Triscone}}, \bibinfo {author} {\bibfnamefont {J.}~\bibnamefont {Junquera}},
  \ and\ \bibinfo {author} {\bibfnamefont {P.}~\bibnamefont {Ghosez}},\ }\href
  {\doibase 10.1103/physrevlett.94.047603} {\bibfield  {journal} {\bibinfo
  {journal} {Phys. Rev. Lett.}\ }\textbf {\bibinfo {volume} {94}},\ \bibinfo
  {pages} {047603} (\bibinfo {year} {2005})}\BibitemShut {NoStop}%
\bibitem [{\citenamefont {Stephenson}\ \emph {et~al.}(2003)\citenamefont
  {Stephenson}, \citenamefont {Fong}, \citenamefont {Murty}, \citenamefont
  {Streiffer}, \citenamefont {Eastman}, \citenamefont {Auciello}, \citenamefont
  {Fuoss}, \citenamefont {Munkholm}, \citenamefont {Aanerud},\ and\
  \citenamefont {Thompson}}]{Stephenson2003}%
  \BibitemOpen
  \bibfield  {author} {\bibinfo {author} {\bibfnamefont {G.}~\bibnamefont
  {Stephenson}}, \bibinfo {author} {\bibfnamefont {D.}~\bibnamefont {Fong}},
  \bibinfo {author} {\bibfnamefont {M.~R.}\ \bibnamefont {Murty}}, \bibinfo
  {author} {\bibfnamefont {S.}~\bibnamefont {Streiffer}}, \bibinfo {author}
  {\bibfnamefont {J.}~\bibnamefont {Eastman}}, \bibinfo {author} {\bibfnamefont
  {O.}~\bibnamefont {Auciello}}, \bibinfo {author} {\bibfnamefont
  {P.}~\bibnamefont {Fuoss}}, \bibinfo {author} {\bibfnamefont
  {A.}~\bibnamefont {Munkholm}}, \bibinfo {author} {\bibfnamefont
  {M.}~\bibnamefont {Aanerud}}, \ and\ \bibinfo {author} {\bibfnamefont
  {C.}~\bibnamefont {Thompson}},\ }\href {\doibase
  10.1016/s0921-4526(03)00273-4} {\bibfield  {journal} {\bibinfo  {journal}
  {Physica B: Condens. Matter}\ }\textbf {\bibinfo {volume} {336}},\ \bibinfo
  {pages} {81} (\bibinfo {year} {2003})}\BibitemShut {NoStop}%
\bibitem [{\citenamefont {Prutton}(1994)}]{prutton1994introduction}%
  \BibitemOpen
  \bibfield  {author} {\bibinfo {author} {\bibfnamefont {M.}~\bibnamefont
  {Prutton}},\ }\href@noop {} {\emph {\bibinfo {title} {Introduction to surface
  physics}}}\ (\bibinfo  {publisher} {Clarendon Press Oxford},\ \bibinfo {year}
  {1994})\ p.~\bibinfo {pages} {58}\BibitemShut {NoStop}%
\bibitem [{\citenamefont {Sun}\ \emph {et~al.}(2019)\citenamefont {Sun},
  \citenamefont {Abid}, \citenamefont {Tan}, \citenamefont {Ren}, \citenamefont
  {Li}, \citenamefont {Li}, \citenamefont {Chen}, \citenamefont {Li},
  \citenamefont {Zhang}, \citenamefont {Zhong} \emph {et~al.}}]{Sun2019}%
  \BibitemOpen
  \bibfield  {author} {\bibinfo {author} {\bibfnamefont {Y.}~\bibnamefont
  {Sun}}, \bibinfo {author} {\bibfnamefont {A.~Y.}\ \bibnamefont {Abid}},
  \bibinfo {author} {\bibfnamefont {C.}~\bibnamefont {Tan}}, \bibinfo {author}
  {\bibfnamefont {C.}~\bibnamefont {Ren}}, \bibinfo {author} {\bibfnamefont
  {M.}~\bibnamefont {Li}}, \bibinfo {author} {\bibfnamefont {N.}~\bibnamefont
  {Li}}, \bibinfo {author} {\bibfnamefont {P.}~\bibnamefont {Chen}}, \bibinfo
  {author} {\bibfnamefont {Y.}~\bibnamefont {Li}}, \bibinfo {author}
  {\bibfnamefont {J.}~\bibnamefont {Zhang}}, \bibinfo {author} {\bibfnamefont
  {X.}~\bibnamefont {Zhong}},  \emph {et~al.},\ }\href {\doibase
  10.1126/sciadv.aav4355} {\bibfield  {journal} {\bibinfo  {journal} {Sci.
  Adv.}\ }\textbf {\bibinfo {volume} {5}},\ \bibinfo {pages} {eaav4355}
  (\bibinfo {year} {2019})}\BibitemShut {NoStop}%
\bibitem [{\citenamefont {Baker}\ and\ \citenamefont
  {Bowler}(2019)}]{Baker2019}%
  \BibitemOpen
  \bibfield  {author} {\bibinfo {author} {\bibfnamefont {J.~S.}\ \bibnamefont
  {Baker}}\ and\ \bibinfo {author} {\bibfnamefont {D.~R.}\ \bibnamefont
  {Bowler}},\ }\href {\doibase 10.1103/physrevb.100.224305} {\bibfield
  {journal} {\bibinfo  {journal} {Phys. Rev. B}\ }\textbf {\bibinfo {volume}
  {100}},\ \bibinfo {pages} {224305} (\bibinfo {year} {2019})}\BibitemShut
  {NoStop}%
\end{thebibliography}%


\begin{thebibliography}{10}

\bibitem{Nakata2014}
A.~Nakata, D.~R. Bowler, and T.~Miyazaki,
  ``\href{https://doi.org/10.1021/ct5004934}{Efficient Calculations with
  Multisite Local Orbitals in a Large-Scale {DFT} Code {CONQUEST}},'' {\em J.
  Chem. Theory Comput}, vol.~10, pp.~4813--4822, oct 2014.

\bibitem{Baker2020}
J.~S. Baker, T.~Miyazaki, and D.~R. Bowler,
  ``\href{https://doi.org/10.1088/2516-1075/ab950e}{The pseudoatomic orbital
  basis: electronic accuracy and soft-mode distortions in ABO$_3$
  perovskites},'' {\em Electron. Struct}, May 2020.

\bibitem{Nakata2020Large}
A.~Nakata, J.~S. Baker, S.~Y. Mujahed, J.~T.~L. Poulton, S.~Arapan, J.~Lin,
  Z.~Raza, S.~Yadav, L.~Truflandier, T.~Miyazaki, and D.~R. Bowler,
  ``\href{https://doi.org/10.1063/5.0005074}{Large scale and linear scaling
  {DFT} with the {CONQUEST} code},'' {\em J. Chem. Phys}, vol.~152, p.~164112,
  Apr. 2020.

\bibitem{CQRelease2020}
D.~R. Bowler, J.~S. Baker, J.~T.~L. Poulton, S.~Y. Mujahed, J.~Lin, S.~Yadav,
  Z.~Raza, T.~Miyazaki, M.~Gillan, A.~Nakata, L.~Truflandier, A.~Torralba,
  V.~Brazdova, L.~Tong, M.~Arita, A.~Sena, U.~Terranova, R.~Choudhury,
  C.~Goringe, E.~Hernandez, and I.~Bush, ``The \textsc{CONQUEST} code: public
  release at
  \href{https://github.com/OrderN/CONQUEST-release}{https://github.com/OrderN/CONQUEST-release},''
  Jan 2020.

\bibitem{bowler2012methods}
D.~Bowler and T.~Miyazaki,
  ``\href{http://iopscience.iop.org/article/10.1088/0034-4885/75/3/036503/meta}{$\mathcal{O}(N)$
  Methods in electronic structure calculations},'' {\em Rep. Prog. Phys},
  vol.~75, no.~3, p.~036503, 2012.

\bibitem{rayson2009highly}
M.~J. Rayson and P.~R. Briddon,
  ``\href{https://doi.org/10.1103/physrevb.80.205104}{Highly efficient method
  for Kohn-Sham density functional calculations of
  500{\textendash}10{\hspace{0.167em}}000 atom systems},'' {\em Phys. Rev. B},
  vol.~80, p.~05104, nov 2009.

\bibitem{Birch1947}
F.~Birch, ``\href{https://doi.org/10.1103/physrev.71.809}{Finite Elastic Strain
  of Cubic Crystals},'' {\em Phys. Rev}, vol.~71, pp.~809--824, June 1947.

\bibitem{Gonze2009}
X.~Gonze, B.~Amadon, P.-M. Anglade, J.-M. Beuken, F.~Bottin, P.~Boulanger,
  F.~Bruneval, D.~Caliste, R.~Caracas, M.~C{\^{o}}t{\'{e}}, {\em et~al.},
  ``\href{https://doi.org/10.1016/j.cpc.2009.07.007}{ABINIT: First-principles
  approach to material and nanosystem properties},'' {\em Comput. Phys.
  Commun}, vol.~180, pp.~2582--2615, Dec. 2009.

\bibitem{Gonze2016}
X.~Gonze, F.~Jollet, F.~A. Araujo, D.~Adams, B.~Amadon, T.~Applencourt,
  C.~Audouze, J.-M. Beuken, J.~Bieder, A.~Bokhanchuk, {\em et~al.},
  ``\href{https://doi.org/10.1016/j.cpc.2016.04.003}{Recent developments in the
  ABINIT software package},'' {\em Comput. Phys. Commun}, vol.~205,
  pp.~106--131, Aug. 2016.

\bibitem{monkhorst1976special}
H.~J. Monkhorst and J.~D. Pack,
  ``\href{https://doi.org/10.1103/physrevb.13.5188}{Special points for
  Brillouin-zone integrations},'' {\em Phys. Rev. B}, vol.~13, pp.~5188--5192,
  jun 1976.

\bibitem{Meyer2002}
B.~Meyer and D.~Vanderbilt,
  ``\href{https://doi.org/10.1103/physrevb.65.104111}{Ab initio study of
  ferroelectric domain walls in PbTiO$_3$},'' {\em Phys. Rev. B}, vol.~65,
  p.~104111, Mar. 2002.

\bibitem{Resta1999}
R.~Resta,
  ``\href{10.1002/(sici)1097-461x(1999)75:4/5<599::aid-qua25>3.0.co;2-8}{Macroscopic
  polarization from electronic wave functions},'' {\em Int. J. Quantum Chem.},
  vol.~75, no.~4-5, pp.~599--606, 1999.

\end{thebibliography}

\end{document}


\author{Jack S. Baker$^{1, 2}$, \& David R. Bowler$^{1, 2, 3}$  \\\small{$^1$\textit{London Centre for Nanotechnology, UCL, 17-19 Gordon St, London WC1H 0AH, UK}} \\ \small{$^2$\textit{Department of Physics \& Astronomy, UCL, Gower St, London WC1E 6BT, UK}} \\ \small{$^3$\textit{International Centre for Materials Nanoarchitectonics (MANA)}} \\ \small{\textit{National Institute for Materials Science (NIMS), 1-1 Namiki, Tsukuba, Ibaraki 305-0044, Japan}}}

\title{\textbf{Supporting Information:} \\ \medskip \Large{Polar morphologies from first principles: PbTiO$_3$ films on SrTiO$_3$ substrates and the $p(2 \times \Lambda)$ surface reconstruction}}
\maketitle

\medskip


This document provides supporting information for the article "\textit{Polar morphologies from first principles: PbTiO$_3$ films on SrTiO$_3$ substrates and the $p(2 \times \Lambda)$ surface reconstruction}". We show here our tests for convergence in the parameters of our simulations. In Section \ref{MSSF}, we show the convergence of the parameters of the multi-site support function (MSSF) method \cite{Nakata2014}. In Section \ref{substrate}, we study the vertical displacement of metal cations as a function of the number of included substrate monolayers. In Section \ref{gridandpack}, we study the convergence of the total energies and energy differences for the fineness of integration grid and Brillouin zone sampling. Lasty, in Section \ref{eq:localpol} we detail the method used for the calculation of local polarization. This method is used to produce the polarization vector field plots in the main text. For details about the generation of the basis sets used in this work we refer the reader to reference \cite{Baker2020} - particularly towards its supplementary material. For further details about the \textsc{CONQUEST} code we refer the reader to a recently written a review article \cite{Nakata2020Large} and to \cite{CQRelease2020} for a public release of the code under an MIT license. If any further information or raw data are required, do not hesitate to contact the authors via the email contacts at the bottom of this page\footnote[2]{Jack S. Baker: \textcolor{blue}{jack.baker.16@ucl.ac.uk}, David R. Bowler: \textcolor{blue}{david.bowler@ucl.ac.uk}}.

\section{Multi-site support function ranges \label{MSSF}}

This work uses the multi-site support function (MSSF) method \cite{Nakata2014} to contract the basis set size and thus the dimensions of the Hamiltonian matrix (along with other matrices). This significantly reduces the computational effort required for its diagonalisation. Since the computational times associated with standard matrix diagonalisation routines scale asymptotically as $\mathcal{O}(N_{\text{basis}}^3)$ \cite{bowler2012methods} (where $N_{\text{basis}}$ is the total number of basis functions) the MSSF contraction is particularly important. This is since the MSSF contraction typically reduces the total number of basis functions in a calculation by a \textit{factor of 3 or 4}. Such savings allow our calculations to venture into the region of $\approx$ 1000-3000 atoms with high accuracy. \par

While we refer to reader to the main text for an explanation of the method (and to references \cite{Nakata2014} and \cite{rayson2009highly}) we recall that the MSSF contraction introduces two new adjustable parameters to the calculation; $r_{MS}$ and $r_{LD}$. We must carefully converge these two parameters to ensure we retain the same accuracy as the uncontracted PAO basis (which we refer to as 'exact diagonalisation'). We do so by calculating the total energy of the high temperature cubic $Pm\bar{3}m$ phases of PTO and STO as a function of volume and fitting the Birch-Murnaghan equation of state \cite{Birch1947}. We also consider the relaxed geometry of the $P4mm$ ferroelectric phase of PTO. \par

\begin{figure}
       \includegraphics[width=\linewidth]{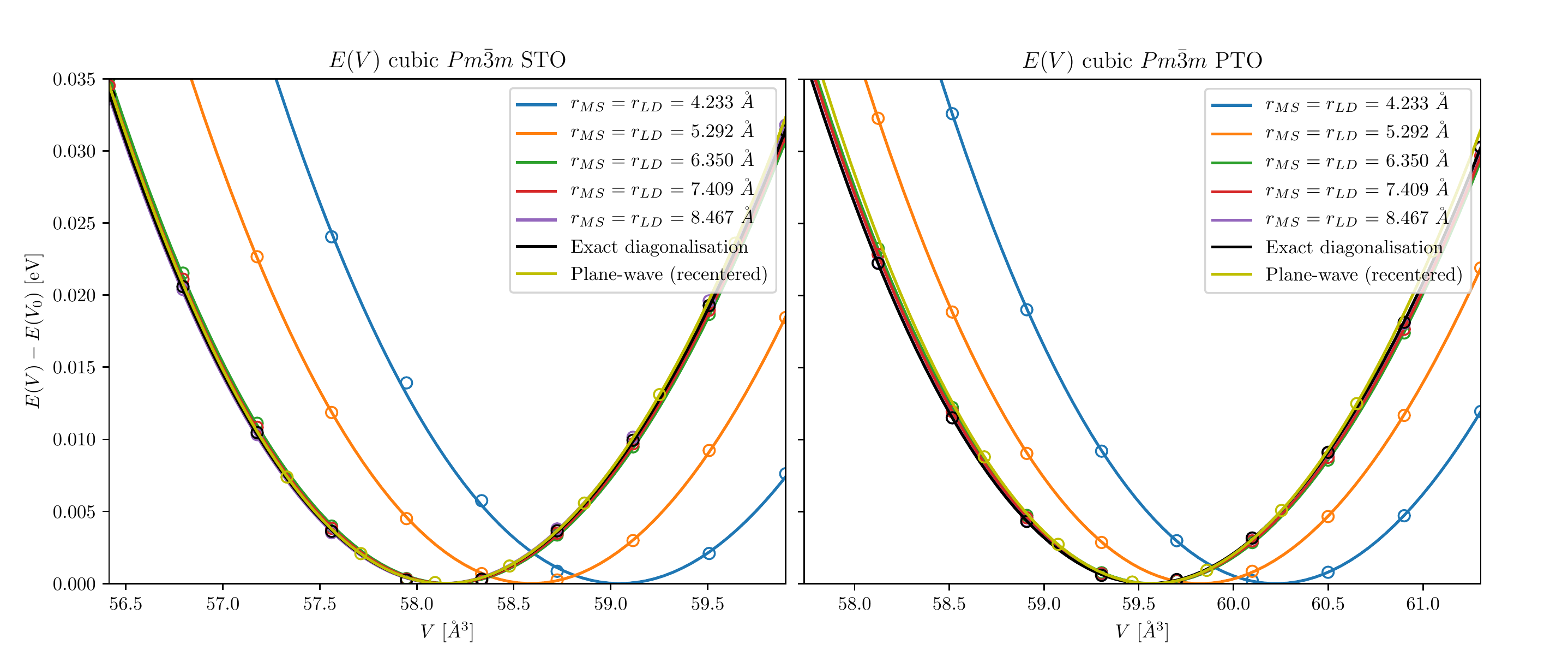}
        \caption{The equations of state $E(V)$ for the cubic phases of STO (left) and PTO (right) using different $r_{MS}=r_{LD}$. We also include $E(V)$ for the exact diagonalistion of PAOs (where the MSSF contraction has not been made) and the $E(V)$ (PW) result obtained from a plane-wave calculation using the same pseudopotentials. The PW curve has $V_0$ shifted to agree with the minimum of the exact diagonalisation calculation so a better comparison can be made of the energetics. The true lattice constants can be found in Table \ref{tab:BMFits}.}
        \label{fig:eos}
    \end{figure}

Figure \ref{fig:eos} shows the equations of state for cubic PTO (Figure \ref{fig:eos} right) and STO (Figure \ref{fig:eos} left) for various $r_{MS} = r_{LD}$ and exact diagonalisation. We also include on each figure the equation of state obtained with a plane-wave (PW) basis set using the same pseudopotential as performed with the \texttt{ABINIT} code (v8.10.3) \cite{Gonze2009, Gonze2016}. Plane-wave and PAO calculations use a $6 \times 6 \times 6$ Monkhorst-Pack mesh for Brillouin zone integrals (convergence is displayed in figure \ref{fig:mpmesh}). Plane-wave calculations use a cut-off energy of 40Ha whilst PAO calculations use a regular integration grid (for wavefunctions and the charge density) with a plane-wave equivalent cutoff of 300Ha (convergence is displayed in figure \ref{fig:intgrid}). We see that by $r_{MS} = r_{LD} = 6.350 \text{\AA}$ excellently agrees with the exact diagonalistation and PW calculation. The parameters of the fit are shown in table \ref{tab:BMFits}. We see once again that by $r_{MS} = r_{LD} = 6.350 \text{\AA}$, all the parameters of the fit closely resemble the exact diagonalisation result. We see also that the lattice constants (calculated from the EOS fit - not lattice vector optimisation which produces a slightly different result) beyond $r_{MS} = r_{LD} = 6.350 \text{\AA}$ agree with the plane-wave calculations to an error of +0.51\% and +0.52\% for PTO and STO respectively. We see also that the Bulk modulus is well described by the PAO basis sets offering errors of -3.55\% and -0.05\% respectively. Finally, we note that all PAO calculations overestimate (by a small amount) the bulk modulus derivative $B_0^{\prime} = (\partial B/ \partial P)_{P = 0}$. This is of little impact however since the Birch-Murnaghan equation of state is not particularly sensitive to $B_0^{\prime}$. Note, for example, that the exact diagonalisation and plane-wave curves of figure \ref{fig:eos} are rather indistinguishable but differ by $\approx 13\%$ in $B_0^{\prime}$. \par

\begin{table}[]
\centering
\begin{tabular}{@{}ccccc@{}}
\toprule \toprule
                       & $V_0$ {[}$\text{\AA}^3${]} & $a_0$ {[}$\text{\AA}${]} & $B_0$ [GPa] & $B_0^{\prime}$ \\ \midrule
\multicolumn{5}{c}{$Pm\bar{3}m$ PTO}                                                                                 \\ \midrule
$r_{MS} = r_{LD} = 4.233 \text{\AA}$  & 60.224                     & 3.920                  & 204.112   & 4.592        \\
$r_{MS} = r_{LD} = 5.292 \text{\AA}$ & 59.824                     & 3.911                  & 201.377   & 5.421        \\
$r_{MS} = r_{LD} = 6.350 \text{\AA}$ & 59.577                     & 3.906                  & 199.333   & 5.371        \\
$r_{MS} = r_{LD} = 7.409 \text{\AA}$ & 59.566                     & 3.905                  & 198.941   & 5.332        \\
$r_{MS} = r_{LD} = 8.467 \text{\AA}$ & 59.549                     & 3.905                  & 198.653   & 5.360        \\
Exact diagonalisation  & 59.547                     & 3.905                  & 198.743   & 5.274        \\
PW                     & 58.616                     & 3.885                  & 206.659   & 4.656        \\ \midrule
\multicolumn{5}{c}{$Pm\bar{3}m$ STO}                                                                                 \\ \midrule
$r_{MS} = r_{LD} = 4.233 \text{\AA}$  & 59.046                     & 3.894                  & 197.298   & 5.518        \\
$r_{MS} = r_{LD} = 5.292 \text{\AA}$ & 58.589                     & 3.884                  & 206.681   & 2.410        \\
$r_{MS} = r_{LD} = 6.350 \text{\AA}$ & 58.168                     & 3.875                  & 201.291   & 4.629        \\
$r_{MS} = r_{LD} = 7.409 \text{\AA}$ & 58.157                     & 3.874                  & 200.872   & 4.558        \\
$r_{MS} = r_{LD} = 8.467 \text{\AA}$ & 58.133                     & 3.874                  & 200.847   & 4.651        \\
Exact diagonalisation  & 58.142                     & 3.874                  & 200.186   & 4.653        \\
PW                     & 57.225                     & 3.854                  & 201.395   & 4.095  \\ \bottomrule \bottomrule     
\end{tabular}
\label{tab:BMFits}
\caption{The values of the equilibrium volumes $V_0$, equilibrium lattice constants $a_0$, bulk moduli $B_0$ and bulk modulus derivatives $B_0^{\prime}$ as obtained from a least squares fit of $E(V)$ to the Birch-Murnaghan equation of state.}
\end{table}

We consider now the relaxed structure of ferroelectric $P4mm$ PTO. We firstly note that the structure of $P4mm$ PTO is unchanged between the exact diagonalisation result and the $r_{MS} = r_{LD} = 6.350 \text{\AA}$ within the resolution of our method. That is, a static calculation using the MSSF contraction on the structure optimised with exact diagonalisation shows that our force tolerance (0.01eV/$\text{\AA}$) is still met. Further, the phase transition energies ($\Delta E = E_{P4mm} - E_{Pm\bar{3}m}$) for the MSSF contracted (-47.38meV) and uncontracted (-47.40meV) basis sets differ only by 0.02meV. The phase transition energy is correct to the plane-wave result (-47.06 meV) to -0.7\% showing that our method is structurally and energetically accurate.

\section{The amount of SrTiO$_3$ substrate \label{substrate}}

\begin{figure}
       \includegraphics[width=\linewidth]{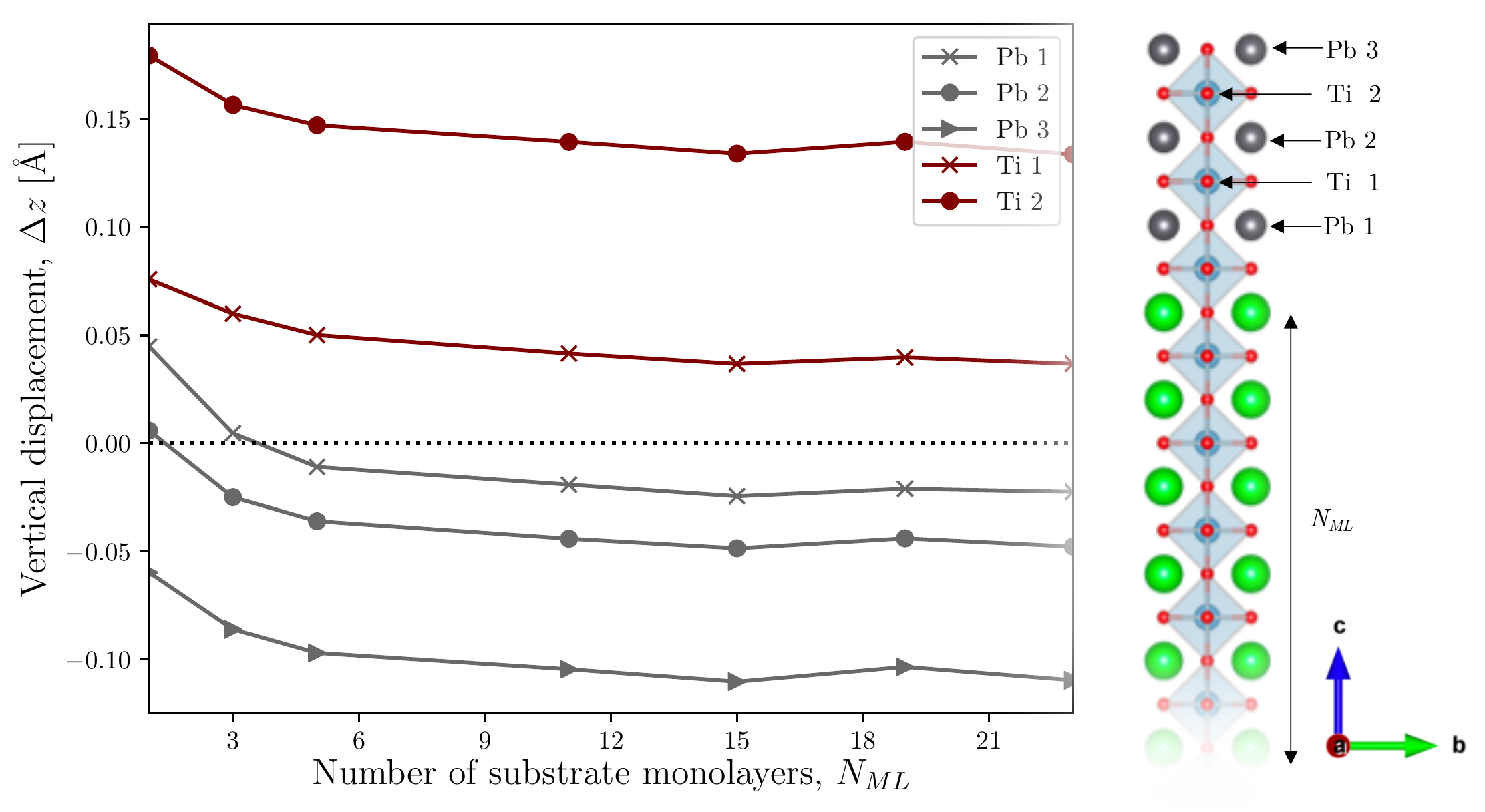}
        \caption{The displacements of metal cations from their initial positions (described in the main text) as a function of the number of included substrate monolayers. }
        \label{fig:substrate}
    \end{figure}

In this section, we present a convergence study for the amount of STO substrate for the simulations in the main text. We perform our tests on the paraelectric phase of the $N_z = 2$ film by including successively more SrO and TiO$_2$ monolayers in the substrate and relaxing each film (with 20 $\text{\AA}$ of vacuum space) to a force tolerance of 0.01 eV/$\text{\AA}$. To two bottom-most monolayers have fixed positions during relaxation to simulate a semi-infinite substrate. We use a uniform integration grid with a plane-wave equivalent cut-off of 300 Ha and a Monhorst-Pack Mesh \cite{monkhorst1976special} of dimensions $6 \times 6 \times 1$ for momentum space integrals. We then measure the vertical displacements $\Delta z$ on the film metal cations from their initial positions and check for convergence as function of the number of substrate monolayers. \par

As can be seen in Figure \ref{fig:substrate}, Pb cations relax towards the substrate whilst Ti Cations relax away from it. This behaviour is only qualitatively achieved beyond $N_{ML} \approx 4$ with the exact positions stabilising beyond $N_{ML} \approx 11$. This suggests that $N_{ML} \geq 11$ should provide a sufficient amount of substrate for our simulations. We do, however, decide to include an additional 4 monolayers ($N_{ML} = 15$) to allow further 'breathing room' for ferroelectric distortions that could penetrate into the substrate when simulating polar phases. Although we find that significant polarization (of the order of the PTO bulk spontaneous polarization) penetrates only about 4-5 monolayers into the substrate, a much smaller background level does penetrate deeper into the substrate. Our inclusion of a larger amount of substrate then allows for this effect.

\section{Integration grids and Monkhorst-Pack meshes \label{gridandpack}}

\begin{figure}
       \includegraphics[width=\linewidth]{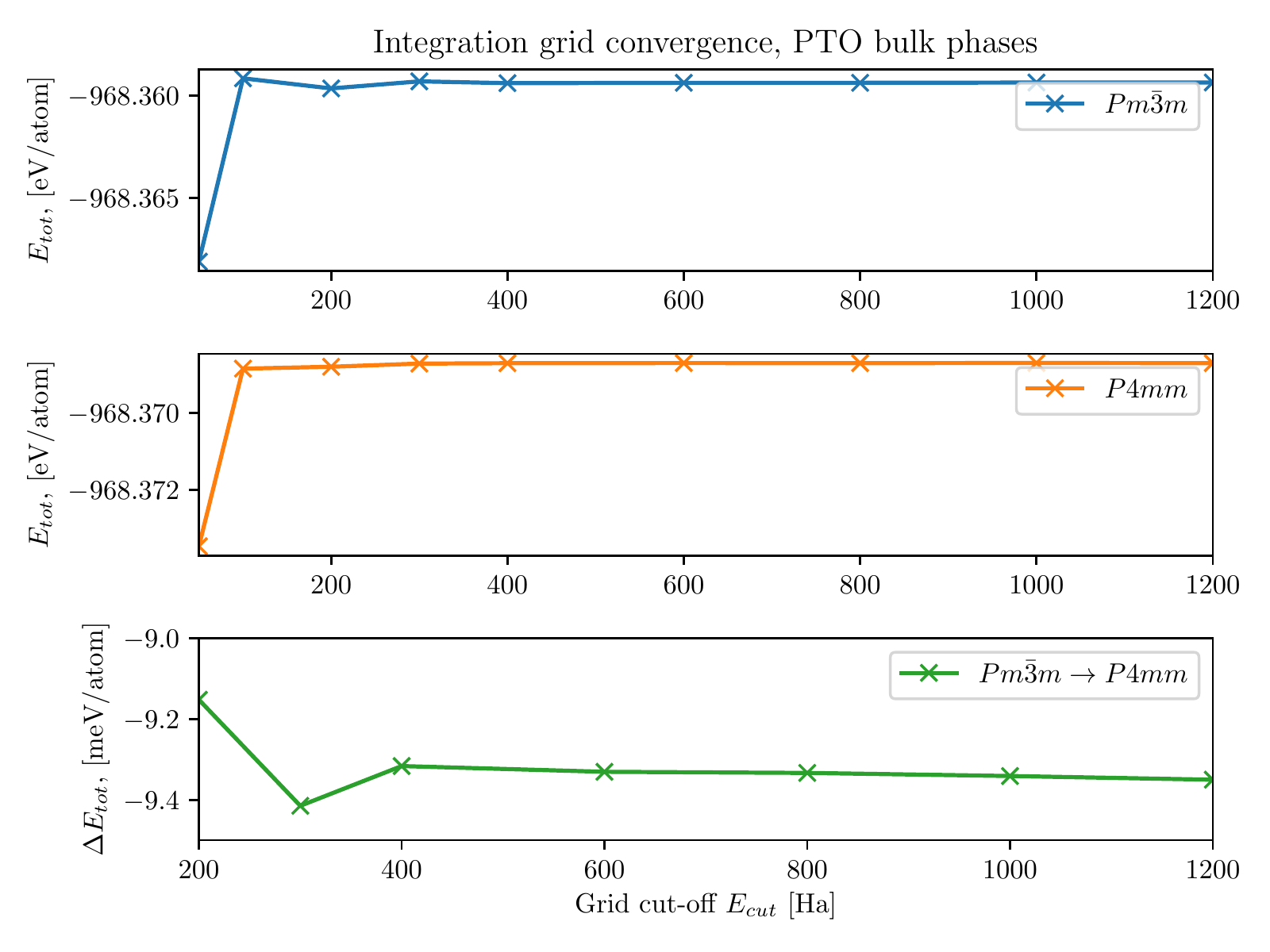}
        \caption{A convergence study for the the fineness of integration grid (measured in a plane-wave equivalent cut-off energy, $E_{cut}$) for the bulk $Pm\bar{3}m$ and $P4mm$ phases of PTO. The upper and middle panels are for convergence of the total energy whilst the lower panel measures the energy difference $\Delta E$ between the phases in meV/atom.}
        \label{fig:intgrid}
    \end{figure}
    
    \begin{figure}
       \includegraphics[width=\linewidth]{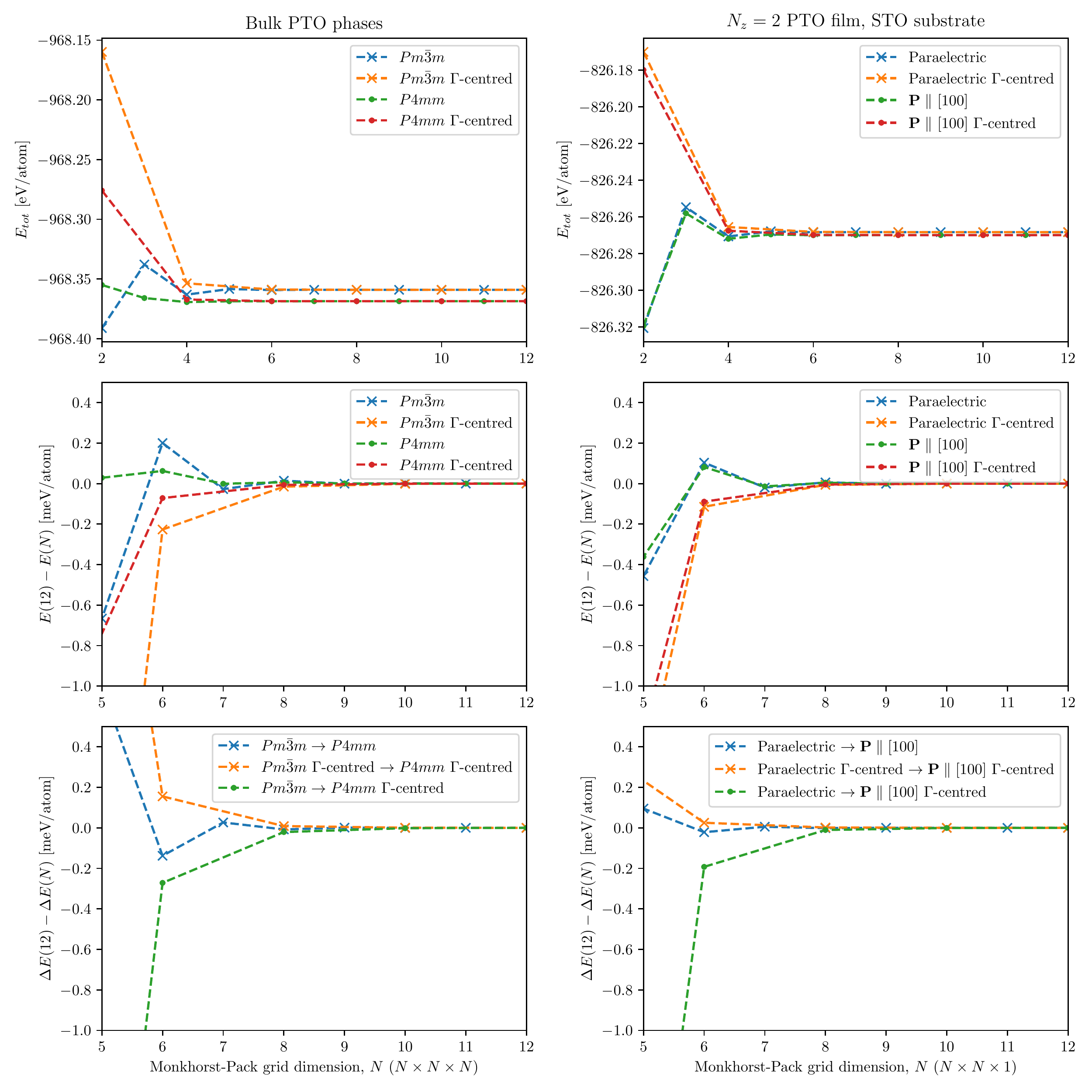}
        \caption{The convergence of reciprocal space integrals using uniform Monkhorst-Pack meshes for the bulk phases of PTO and the $N_z = 2$ film (detailed in the main article). The upper panels show the convergence of the total energy with increasing grid dimensions. The middle panel shows the same test but as a difference with the highly converged (to $\approx 5\times 10^{-7}$ eV in the total energy) $12 \times 12 \times 12$ mesh. The lower panel also compares with the $12 \times 12 \times 12$ mesh, but measures the energy difference between the considered phases.}
        \label{fig:mpmesh}
    \end{figure}

In this section we show convergence in the total energy (per atom, to be consistent with the main text) and energy differences for increasingly fine real-space integration grids \cite{bowler2012methods} and momentum-space Monkhorst-pack \cite{monkhorst1976special} meshes. Figure \ref{fig:intgrid} shows energy convergence for the fineness of integration grid for the bulk phases of PTO. We see that our choice of a 300Ha cut-off is correct to the (very fine) 1200Ha cut-off grid to 0.02 meV/atom in the energy difference (lower panel).

Figure \ref{fig:mpmesh} shows the convergence in the total energy and energy differences with respect to the fineness of Monkhorst-Pack mesh for the bulk phases of PTO (left column) and the $N_z = 2$ film (right column, geometries explained in the main text). It is interesting to point out that energy convergence for the film geometry (a 2D mesh of $N \times N \times 1$) is achieved at a coarser mesh than the bulk geometry (a 3D mesh of $N \times N \times N$). We suggest that this is because convergence in the the out-of-plane direction for the film geometry is (close to) perfectly achieved with a single $\mathbf{k}$-point as ensured by the large vacuum region separating periodic images of the film. We note that there is a small effect related to energy differences between $\Gamma$-centered and non-$\Gamma$-centred meshes. For the film geometry, wee see that the energy difference is lower by 0.19meV/atom for the $6 \times 6\times 1$ mesh (Figure \ref{fig:mpmesh}, bottom right) when one phase is $\Gamma$-centred and the other is not. This effect reduces to a difference of 0.1meV/atom if we consider the energy difference between a paraelectric film (not centred on $\Gamma$) and a film with AFD modes (which is centered on $\Gamma$ since we use a  $6/N_x\times 6/N_y\times 1$ mesh and $N_x = N_y = 2$). We therefore conclude that energy differences between geometries without AFD modes and with AFD modes will all be lower in energy by 0.1meV/atom. This extra degree of energy lowering should be taken into accoutn when examining Figure 3 in the main text, but, makes no difference to the conclusions drawn in the main text (i.e, the favorability hierarchy of different phases). 

\section{\label{local P} The local polarization}

    We define the local polarization $\mathbf{P}^{(i)}$ in the manner suggested by Meyer and Vanderbilt \cite{Meyer2002}
    
    \begin{equation}
        \mathbf{P}^{(i)} = \frac{e}{\Omega_c} \sum_{\alpha} w_{\alpha} \mathbf{Z}_{\alpha}^* \cdot \mathbf{u}_{\alpha}^{(i)}
        \label{eq:localpol}
    \end{equation}
    
    for local unit cell $i$, local unit cell volume $\Omega_c$, atom index $\alpha$, Born effective charge tensor $\mathbf{Z}_{\alpha}^*$, local atomic displacements from the idealised \textit{cubic} bulk positions $\mathbf{u}_{\alpha}^{(i)}$ and atomic weight factor $w_\alpha$ which is related to how many atoms $\alpha$ belong to the local cell $i$. We note that whilst the concept of a local polarization itself is rather troublesome (since the electrical polarization is a macroscopic quantity), equation \ref{eq:localpol} is a useful probe of the local mode and the resulting dipole moment. We must carefully note, however, that there are indeed several choices for the local unit cell $i$. These choices are discussed in the work of Meyer and Vanderbilt \cite{Meyer2002}. We chose to work with two definitions, the Ti-centered unit cell (where Pb or Sr is at the origin) and the Pb/Sr-centered unit cell (where Ti is at the origin). This allows for the calculation of local polarization vectors centered at the both the metal cation sites. For the Ti-centered cells, we have the atomic weight factors $w_{\text{Ti}}=1$, $w_{\text{Pb}}=1/8$ and $w_{\text{O}}=1/2$. For the Pb-centered cells we have $w_{\text{Ti}}=1/8$, $w_{\text{Pb}}=1$ and $w_{\text{O}}=1/2$.
    
    The Born effective charge tensors were calculated for cubic bulk PTO and STO using finite differences in the macroscopic polarization. We do so by displacing each of the symmetry inequivalent sites within a $5 \times 5 \times 5$ supercell (625 atoms) of PTO and STO a small amount (0.005 $\text{\AA}$) and calculating the resulting polarization with Resta's method \cite{Resta1999}. For consistency, we use the $r_{\text{MS}} = r_{\text{LD}} = 6.35 \text{\AA}$ MSSF contraction in the calculation. For PTO, the 3x3 symmetrical diagonal tensor has elements $\mathbf{Z}^*_{\text{Pb, PTO}} = 3.89$ and $\mathbf{Z}^*_{\text{Ti, PTO}} = 7.08$ whilst Oxygen is
    
    \[
    \mathbf{Z}^*_{\text{O, PTO}}=
      \begin{bmatrix}
        -5.79 & 0 & 0  \\
        0 & -2.58 & 0  \\
        0 & 0 & -2.58
      \end{bmatrix}
     \] 
     
     for STO, the 3x3 symmetrical diagonal tensor has elements $\mathbf{Z}^*_{\text{Sr, STO}} = 2.55$ and $\mathbf{Z}^*_{\text{Ti, STO}} = 7.17$ whilst Oxygen is
     
     \[
    \mathbf{Z}^*_{\text{O, STO}}=
      \begin{bmatrix}
        -4.95 & 0 & 0  \\
        0 & -1.92 & 0  \\
        0 & 0 & -1.92
      \end{bmatrix}
     \] 
    
    We note also the the elements of the Oxygen tensors reorder based on which site in the local unit cell we are considering. A drawback of this method is that a local unit cell cannot always be found. That is, at the film surface, there will always be an incomplete unit cell. For this reason, our calculations always finish with a vector centered on Ti and \textit{not} Pb, despite treating the PbO termination. At the PTO/STO interface, there is also a hybrid unit cell with half its A-sites being Pb and half being Sr. For this area, we define no local polarization vectors.

\bibliographystyle{ieeetr}
\bibliography{supplement.bib}